\documentclass[%
reprint,
 amsmath,amssymb,
 aps,
floatfix,
]{revtex4-1}

\usepackage{graphicx}
\usepackage{dcolumn}
\usepackage{amsmath}
\usepackage{amssymb}
\usepackage{bm}
\usepackage[bbgreekl]{mathbbol}
\usepackage{color} 
\usepackage{braket}
\usepackage{gensymb}
\usepackage[dvipsnames]{xcolor}
\usepackage{lpic}
\usepackage[percent]{overpic}
\usepackage{stackengine}
\usepackage{physics}
\usepackage{siunitx}
\usepackage{textcomp}
\usepackage{slashbox}
\usepackage{lipsum}

\usepackage[mathscr]{eucal}

\usepackage{amsfonts}
\usepackage[colorinlistoftodos]{todonotes}
\usepackage{mathtools}
\usepackage{empheq}
\usepackage{esvect}
\usepackage{rotating}

\DeclareSymbolFontAlphabet{\mathbbm}{bbold}
\DeclareSymbolFontAlphabet{\mathbb}{AMSb}

\begin{document}

\preprint{APS/123-QED}

\title{Floquet description of Optically Pumped Magnetometers}

\author{Hans Marin Florez$^{1}$}
\email{hans@if.usp.br}
\author{Tadas Pyragius$^2$}
\email{tadas.pyragius@nottingham.ac.uk}
\affiliation{
$^{1}$Instituto de F\'{\i}sica, Universidade de S\~ao Paulo, 05315-970 S\~ao Paulo, SP-Brazil\\
$^{2}$School of Physics \& Astronomy, University of Nottingham, University Park, Nottingham NG7 2RD, UK
}

\date{\today}

\begin{abstract}
We present theoretical description of Voigt and Faraday effect based optically pumped magnetometers using the Floquet expansion. Our analysis describes the spin-operator dynamics of the first, $\hat{F}(t)$, and second, $\hat{F}^2(t)$, order moments and takes into account of different pumping profiles and decoherence effects. 
We find that the theoretical results are in good agreement with the experimental demonstrations over a wide range of fields and pumping conditions. Finally, the theoretical analysis presented here is generalized and can be extended to different magnetometry schemes with arbitrary pumping profiles and multiple radio-frequency fields.
\end{abstract}

\pacs{Valid PACS appear here}
\maketitle

\section{\label{sec:intro}Introduction}

Atomic vapour based optically pumped magnetometers (OPMs)~\cite{OPM1,OPM2} have become state-of-the-art magnetic field sensors with numerous applications in very diverse areas, ranging from fundamental physics in searching for electric dipole moment (EDM)~\cite{edm1,edm2}, to geophysical and space magnetometry, medicine, such as magneto-encephalography (MEG)~\cite{meg1,meg2} and magneto-cardiography~\cite{mcg0,mcg1,mcg2}.
A number of different OPM architectures have shown sensitivity of fT$/\sqrt{\mathrm{Hz}}$, based on spin-exchange relaxation-free (SERF) magnetometers~\cite{Romalis2002,Romalis2010}, radio-frequency (rf) excited spin with $M_x$ and $M_z$ magnetometers relying on a linear atomic response~\cite{Bison03,kasper10,witold12}, and  modulated light magnetometers producing nonlinear magneto-optical rotation (NMOR)~\cite{budker2002,Gawlik2006} due to a nonlinear optical response of the atoms.

\begin{figure}[bt!]\hspace*{-0.7cm}
\begin{overpic}[width=8.9cm]{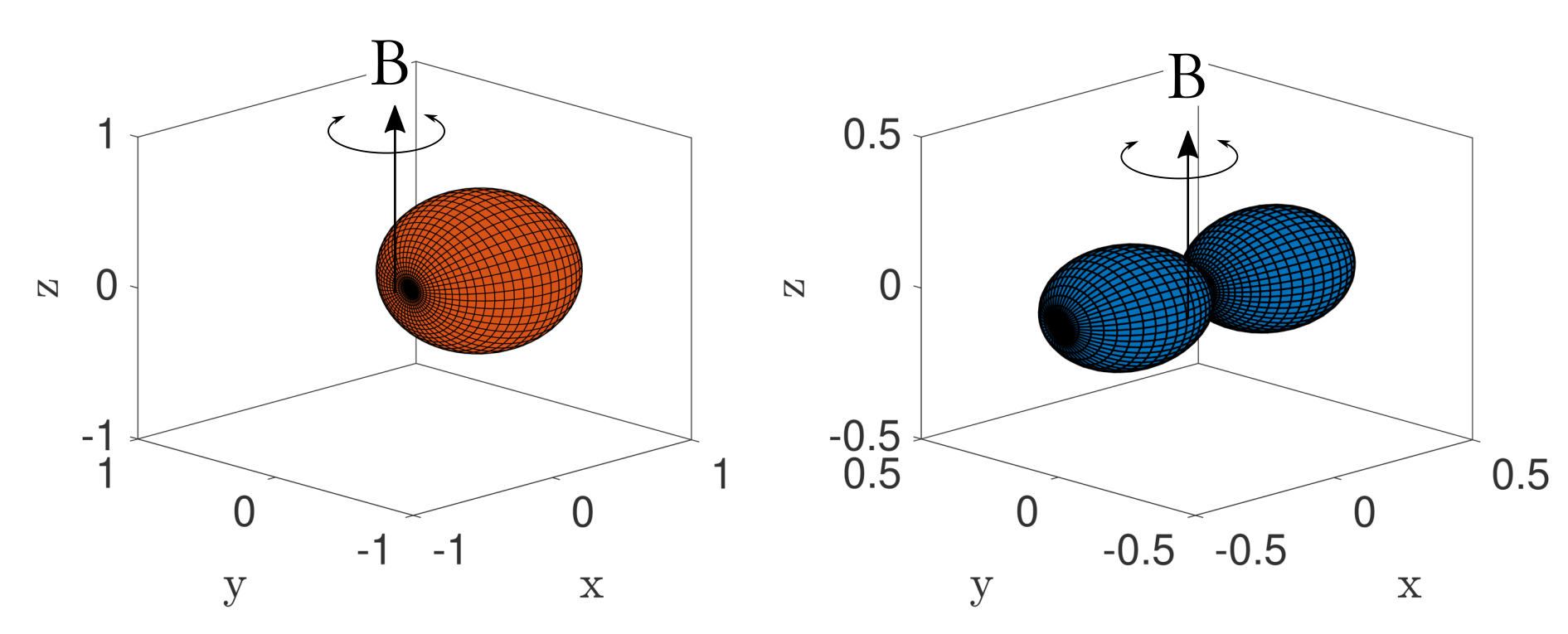}
 \put (5,35) {(a)} 
  \put (50,35) {(b)} 
\end{overpic}
\caption{State angular momentum probability surfaces. (a) Probability surface for an oriented state  $\Ket{F=2,m_F=+2}_x$ state i.e. states with preferred directions. (b) Probability surface for a statistical mixture of $\Ket{F=2,m_F=\pm 2}_x$ states, which correspond to an aligned state i.e. states with preferred axis but no preferred directions. Both states presses when a magnetic field is applied in the $\hat{z}$ direction.}\label{fig:surfaces}
\end{figure}

\begin{figure}[t!]
\begin{overpic}[width=8.9cm]{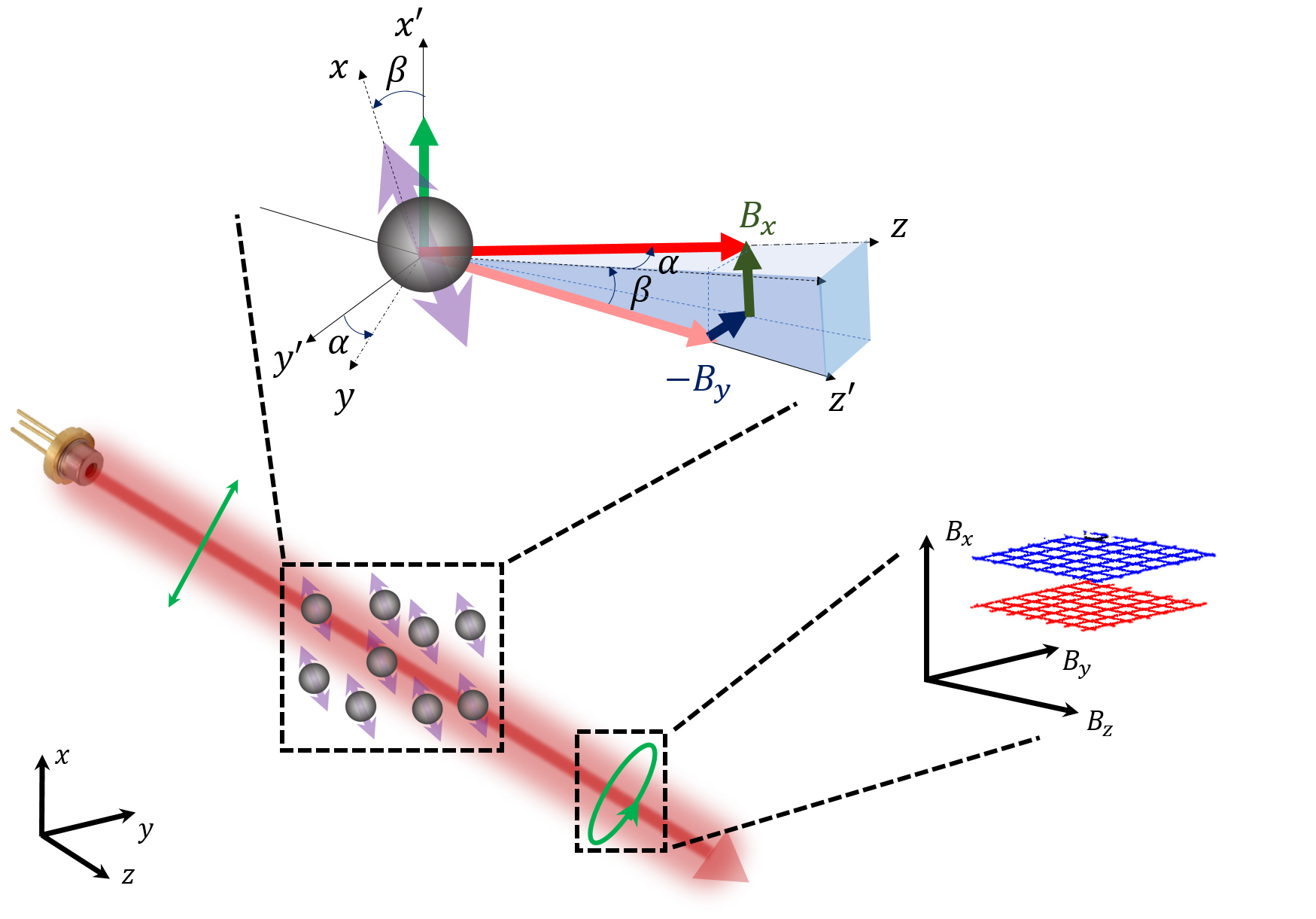}
\end{overpic}
\caption{The laser interacts with the atomic ensemble allowing a dispersive measurement of the aligned state dynamics. The dynamical evolution of aligned states dressed by a radio-frequency field enables detection of all three vector components of the magnetic fields. The magnetic fields $B_x,\ B_y$ and $B_z$ are measured from the change in ellipticity on the probe beam 
and can be represented as a frame rotation from $(x',y',z')$ to $(x,y,z)$\cite{Tadas19}.
}\label{fig:illustration}
\end{figure}

Most of OPMs are based on a Faraday dispersive measurement, in which oriented states (see Fig.~\ref{fig:surfaces}~(a)) are prepared and probed by a detuned laser beam measuring the Faraday rotation induced by the spin polarized sample. This kind of configuration can run in scalar or vector mode~\cite{Romalis2004,Gao2016}, typically in orthogonal geometry.
A  different approach has been shown in ref.~\cite{Weis06}, in which an aligned state is prepared instead of an oriented one (see for example Fig.~\ref{fig:surfaces}~(b)) and read through paramagnetic resonance i.e. non-dispersive measurement. This kind of state allows a vector magnetometer operation using radio-frequency fields~\cite{ingleby17},  or, as it was proposed more recently~\cite{LeGal19}, adopting an all-optical approach which performs a dual axis magnetometer based on Hanle effect. On the other hand, in ref.~\cite{Tadas19} we have shown that indeed it is possible to employ dispersive measurements based on  Voigt rotation when working with aligned states driven by radio frequency fields, also showing vector magnetometry operation, (see Fig.~\ref{fig:illustration}). However, the typical description for oriented spins probed by the Faraday rotation, which depends on the first moment of the spin operator is not suitable for the aligned states probing the Voigt rotation as the latter is proportional to second moment spin operator.

Another important and common feature of the oriented and aligned spin state magnetometers is the use of synchronous pumping in which the amplitude (or frequency) modulation avoids optical decoherence due to the pump. This implies that the pumping rate and the decoherence rate 
(e.g. square waves intensity profile) are in general time dependent.
This type of time modulation leads to interaction with many harmonics, which is typically avoided, by assuming that the pumping is weak, or approximating the interaction by neglecting higher harmonic terms.
In this work, we show an approach that is capable of encompassing these previously neglected time dependent terms. To do so, we employ a Floquet expansion to solve the spin dynamics which enables us to build a more realistic picture of the spin evolution~\cite{Levante95,Bain01}.
The spectral decomposition of the Floquet expansion allow us to address the solutions for multiple harmonics generated in the dynamics independently, which can be directly compared with the experimental observations.

In this paper we present the dynamics of the first and second moment elements when the spins are driven by a radio frequency field, which describes the dynamics of oriented and aligned states. Furthermore, we find the solution for the spin dynamics in the realistic situation where the optical pumping presents an arbitrary time dependence. We show that the general dynamics of the second moment in the Liouville space can be reduced to Bloch equations (see eq.~(\ref{eq:dPdtClassical})) and can be solved by employing the Floquet expansion, which to our knowledge has not been reported before. This solution predicts sensitivity to all three vector components of the magnetic field as reported in the experimental work in \cite{Tadas19}. This approach is also compatible with scenarios in which multiple radio-frequency fields are used.

The paper is organized as follows. In Section~\ref{Sec:Dispersive_Measurements} we introduce the differences between the Faraday and Voigt rotation in the dispersive regime with respect to the statistical moments of the spin operators. 
In Section~\ref{Sec:SpinDynamics}, we study the spin dynamics including the interaction with static and  radio-frequency magnetic fields, and  consider an arbitrary time dependence for the optical pumping. Section~\ref{Sec:FirstMoment} describes the Floquet expansion to solve the dynamics of the first moment. Section~\ref{Sec:2ndtMoment}  describes the dynamics of the second moment of the spins and the transformation to the Liouville space. Section~\ref{Sec:Floquet2ndMoment} shows the results on the Floquet expansion to solve the dynamics and we discuss the main features of the model. Section~\ref{sec:Conclusions} presents our conclusions.

\section{Dispersive optical measurements\label{Sec:Dispersive_Measurements}}

The most common configuration for linear optical devices uses dispersive optical measurements of spin-polarized atoms to detect the presence of an external magnetic field.
The modulation of the birefringence of the medium caused by the Larmor precession of such spin-polarized atoms can be observed polarimetrically, this is known as Faraday rotation. 
More specifically, in terms of polarization moments, the spin oriented atoms represented by the probability surface in Fig.~\ref{fig:surfaces}~(a), for which the z-component of the total angular momentum $\hat{\mathbf{F}}=(\hat{F}_x,\hat{F}_y,\hat{F}_z)$ on average satisfies $\Braket{\hat{F}_z(t)}\neq0$, 
induces a polarization rotation of an incident linearly polarized light that propagates along the $z$-direction \cite{polarised_atoms, rochester}.
Assuming that such interaction is with an atomically thin sample with no back action effects, the rotation can described by the Stokes' parameter. For linearly polarized such light-matter interaction is given by
\begin{align}
\Braket{\hat{S}_x'(t)}&=\Braket{\hat{S}_x(t)}-
G_{F}^{(1)} S_y n_F \Braket{\hat{F}_z(t)},\label{eq:Faraday}
\end{align}
where the prime indicates the output field after interacting with the atomic medium, $\hat{S}_z=(c/2)(\hat{a}^{\dagger}_{+}\hat{a}_{+}-\hat{a}^{\dagger}_{-}\hat{a}_{-})$ and $\hat{S}_y=(c/2)(i\hat{a}^{\dagger}_{-}\hat{a}_{+}-i\hat{a}^{\dagger}_{+}\hat{a}_{-})$ represent the photon flux of elliptical and at $45\degree$ polarized light expressed in terms of creation and annihilation operators $\hat{a}_{\pm}$ and $\hat{a}_{\pm}^\dagger$ for circular polarization components;  $G_{F}^{(k)}$ is the rank-k coupling strength and $n_F$ are the atoms with the same $F$-manifold state \cite{Jammi18}. 

In terms of statistical definitions, the average value of the spin operator $\Braket{\hat{F}_i(t)}$ where $i=x,y$ and $z$, correspond to the first moment of a statistical distribution of the spin operator $\hat{F}_i(t)$. Thus, we can claim that OPMs based on the dispersive Faraday rotation work only for quantum states with non-zero first moments. Since $\Braket{\hat{F}_i(t)}$ correspond to the classical description of polarized samples, the dynamics of OPMs based on rf excitation or optical excitation, are classically described by the Bloch equations \cite{Bloch46} 
\begin{align}
    \frac{d \mathbf{P}}{dt}=\gamma \mathbf{B}\times \mathbf{P} -\frac{1}{T_2}(P_x \mathbf{e}_x+P_y \mathbf{e}_y)-\frac{1}{T_1}(P_z-P_0)\mathbf{e}_z,\label{eq:dPdtClassical}
\end{align}
where $\gamma$ is proportional to the Larmor frequency and  the polarization vector is $ \mathbf{P}=(P_x,P_y,P_z)$ with $P_i=\Braket{\hat{F}_i}$. The decoherence $T_1$ accounts for relaxation of the spins along the longitudinal direction from magnetic field gradients,  collisions with the walls and atoms that are pumped with spin polarization $P_0$.
The decoherence of the transverse polarization is described by a term proportional to $1/T_2$, and represents the atom-atom collisions.
When the spins are driven by a sinusoidal magnetic field $\mathbf{B}$ the spin dynamics follows a resonance response, which is given by the Bloch solution \cite{Bloch46,Seltzer08}. 


We can also consider the precession of aligned states as the one represented in Fig.~\ref{fig:surfaces}~(b), around a static magnetic field. 
To optically probe this kind of dynamics dispersively, we have proposed in ref.~\cite{Tadas19} a measurement based on Voigt rotation. This effect measures the changes in the linear birefringence of the probing light  (see Fig.~\ref{fig:illustration}). In the limit of a far detuned probe light traversing an atomically thin sample and assuming no back-action effects, the changes in the linear birefringence can be described by the following Stokes' parameters \cite{Julsgaard2003}
\begin{align}
\Braket{\hat{S}_z'(t)}&=\Braket{\hat{S}_z(t)}+
G_{F}^{(2)} S_y n_F \Braket{\hat{F}_x^2(t)-\hat{F}_y^2(t)}\label{eq:Voigt},
\end{align}
which is proportional to transverse second moments of the spin operators i.e the average of the second order products $\Braket{\hat{F}_i(t)\hat{F}_j(t)}$ with $i,j=x,y$ and $z$. Unfortunately, the dynamics for the second moments cannot be described by the Bloch equations, eq.~(\ref{eq:dPdtClassical}). The dynamics for the first moment $P_i=\langle \hat{F}_i\rangle$, which is the average of a linear operator,  in general is different from the dynamics for the second moments $\Braket{\hat{F}_i(t)\hat{F}_j(t)}$, which is the average of bi-linear operators. In particular, there is no Faraday rotation for an aligned state out of the oriented-to-alignment conversion (OAC) regime, which yields a trivial solution for the Bloch equation.
Hence, it would be desirable to have a  dynamical equation like eq.~(\ref{eq:dPdtClassical}), but for the second moment operators based on Heisenberg equations of motion.


\section{Spin dynamics in a radio-frequency dressed field with optical pumping and relaxation\label{Sec:SpinDynamics}}
\subsection{Heisenberg-Langevin equations}
Consider a radio frequency dressing field in the presence of transverse and longitudinal magnetic fields
\begin{equation}
\mathbf{B}=(B_{\mathrm{rf}} \cos{\omega t} +B_x^{\mathrm{ext}})\mathbf{e}_x+B_y^{\mathrm{ext}}\mathbf{e}_y+(B_{\mathrm{dc}}+B_z^{\mathrm{ext}})\mathbf{e}_z,
\end{equation}
where $B_{\mathrm{rf}}$ is the amplitude of the rf field and $B_{\mathrm{dc}}$ is the static field along the longitudinal direction, which are experimentally controlled. Additionally, we have the external fields
$B_i^{\mathrm{ext}}$ with $i=x,y$ and $z$, which originate from external sources. For small fields, where the second order Zeeman shift can be neglected, 
the magnetic field interaction is given by $\hat{H}=(\mu_Bg_F/\hbar)\hat{\mathbf{F}}\cdot\mathbf{B}$, such that in the Heisenberg picture we have
\begin{align}
\hat{H}(t)&=(\Omega_{\mathrm{rf}}\cos(\omega t)+\Omega_{x}^{\mathrm{ext}})\hat{F}_x(t)\nonumber\\
& + \Omega_y^{\mathrm{ext}}\hat{F}_y(t)+(\Omega_{\mathrm{dc}}+\Omega_z^{\mathrm{ext}})\hat{F}_z(t),
\end{align}
with $g'_F=g_F/\hbar$ and $\Omega_i=\mu_B g'_F B_i$ with $i=\mathrm{rf},\mathrm{dc},x,y$ and $z$.
The coherent part of the atomic spin dynamics is given by
the Heisenberg equation $\left.\frac{d\hat{F}_i(t)}{dt}\right|_{\mathrm{coher}}=-\frac{i}{\hbar}[\hat{F}_i(t),\hat{H}(t)]$
such that the spin dynamics due to the magnetic fields is given by

\begin{align}
\left.\frac{d\mathbf{\hat{F}}(t)}{dt}\right|_{\textrm{coher}}&=(\mathbf{B}_0(t)+ \mathbf{B}^{\mathrm{ext}}) \mathbf{\hat{F}}(t),
\end{align}
where the atomic spin vector is  $\mathbf{\hat{F}}(t)=(\hat{F}_x(t),\hat{F}_y(t),\hat{F}_z(t))$ and 
\begin{align}
\mathbf{B}_0(t)&=\begin{bmatrix}
    0 & -\Omega_{\mathrm{dc}} & 0 \\
    \Omega_{\mathrm{dc}} & 0 & -\Omega_{\mathrm{rf}}\cos(\omega t) \\
    0 & \Omega_{\mathrm{rf}}\cos(\omega t) & 0
\end{bmatrix}, \\
\mathbf{B}^{\mathrm{ext}}&=\begin{bmatrix}
    0 & -\Omega_z^{\mathrm{ext}} & \Omega_y^{\mathrm{ext}} \\
    \Omega_z^{\mathrm{ext}} & 0 & -\Omega_x^{\mathrm{ext}} \\
   -\Omega_y^{\mathrm{ext}}+ & \Omega_x^{\mathrm{ext}} & 0
\end{bmatrix}.
\end{align}
Now, to describe the full dynamics of the magnetometer, we need to include additional terms, which govern the pumping and the decay processes of the prepared spin states. One of the contributions corresponding to the state preparation process is the pumping term, namely
\begin{align}
\left.\frac{d\mathbf{\hat{F}}(t)}{dt}\right|_{\Gamma_p}&=-\Gamma_p(t) \mathbf{\hat{F}}(t)+\mathbf{\hat{F}}^{\textrm{in}}(t),
\end{align}
where the pump rate $\Gamma_p(t)$ describes a general form in time at which the state preparation is done e.g. synchronous pumping with any harmonic profile. We will later describe the harmonic decomposition of $\Gamma_p(t)$. We consider the action of the pump process as stochastic flips in time on the atomic operator. Therefore, in general, we consider $\mathbf{\hat{F}}^{\textrm{in}}(t)$ as a stochastic vector operator with non-zero mean value. 
Thus, we propose a linearized-like version of the input operator i.e. $\hat{O}=\Braket{O}+\delta\hat{O}$, such that
 \begin{align}
\mathbf{\hat{F}}^{\textrm{in}}(t)=\Gamma_p(t)\Braket{\mathbf{\hat{F}}^{\textrm{in}}}+\bm{\mathcal{\hat{F}}}^{\textrm{in}}(t),\label{eq:StochFin}
\end{align}
with non-zero mean value, where the stochastic part satisfies $\Braket{\bm{\mathcal{\hat{F}}}^{\textrm{in}}(t)}=0$ and 
its correlation function is
\begin{align}
\Braket{ \bm{\mathcal{\hat{F}}}^{\textrm{in}}(t) \bm{\mathcal{\hat{F}}}^{\textrm{in}}(t)^T}=\Gamma_p(t)\bm{\sigma}_{\textrm{in}}\ \delta(t-t'),\label{eq:FinFinStoch}
\end{align}
where $\bm{\sigma}_{\textrm{in}}$ is the input second moment matrix, which acts as a diffusion term in the second moment dynamics. 
This input operator that we propose recovers the pumping term in the Bloch eq.~(\ref{eq:dPdtClassical}), since $\Braket{\mathbf{\hat{F}}^{\textrm{in}}}=P_0\mathbf{e}_z$.  By adopting this linearized version of the input operator, which follows a perturbative approach of the pump action into spin dynamics, splits the contributions given by the first moment and the second moments of the input state.
The second term we want to include is related to the relaxation of the spins. We apply a relaxation process in terms of stochastic operators $\bm{\mathcal{\hat{F}}}(t)=(\mathcal{\hat{F}}_x(t),\mathcal{\hat{F}}_y(t),\mathcal{\hat{F}}_z(t))$, to obtain the Langevin dynamics of the atomic spin operators
\begin{align}
\left.\frac{d\mathbf{\hat{F}}(t)}{dt}\right|_{\Gamma_\mathrm{rel}}&= - \mathbf{\Gamma}_\mathrm{rel} \mathbf{\hat{F}}(t)+\bm{\mathcal{\hat{F}}}(t),
\end{align}
where $\Braket{\bm{\mathcal{\hat{F}}}(t)}=0$ and in the general case the relaxation matrix $\mathbf{\Gamma}_\mathrm{rel}$ is a diagonal matrix with components $\Gamma_i$ with $i=x,y$ and $z$.
The stochastic operator associated to this relaxation process satisfies the correlation function
\begin{align}
\Braket{\bm{\mathcal{\hat{F}}}_i(t) \bm{\mathcal{\hat{F}}}_j(t')}=(\tilde{\mathbf{\Gamma}})_{ij}\delta(t-t'),\label{eq:FFstochasticC}
\end{align}
where $\tilde{\mathbf{\Gamma}}$ correspond to the diffusion matrix.
In Appendix \ref{App:DiffusionM} we formulate the diffusion matrix in terms of the relaxation matrix $\mathbf{\Gamma}_\mathrm{rel}$ satisfying operator commutation relations.

In general, each direction is subjected to different decoherence rates. Nevertheless, 
the relaxation processes in the transverse directions are typically different from the longitudinal direction. As a result, it is commonly considered that the transverse direction is affected equally by spin exchange collisions such that $\Gamma_x=\Gamma_y=\Gamma_2$. On the other hand, processes like wall collision, decoherence induced by the pump and magnetic field gradients, may relax the longitudinal direction at a different rate, given by $\Gamma_z=\Gamma_1$.

Combining the terms containing coherent spin dynamics, pumping and relaxation, we obtain the total spin dynamics of the system 
\begin{align}
\frac{d\mathbf{\hat{F}}(t)}{dt}=&\left(\mathbf{B}_0(t)+ \mathbf{B}_{\mathrm{ext}}^{(0)}\right) \mathbf{\hat{F}}(t)-\Gamma_\mathrm{rel}\mathbf{\hat{F}}(t)
-\Gamma_p(t)\mathbf{\hat{F}}(t)\nonumber\\
&+ \mathbf{\hat{F}}^{\textrm{in}}(t)+\bm{\mathcal{\hat{F}}}(t).\label{eq:FulldfdtLabStoch}
\end{align}

\section{Dynamics of the first moment $\Braket{\mathbf{\hat{F}}'(t)}$\label{Sec:FirstMoment} }
\subsection{Spin dynamics in the laboratory frame \label{sec:LabFrame}}
The dynamics of the first moment is defined by the mean value of eq.~(\ref{eq:FulldfdtLabStoch}), such that $\mathbf{P}(t)=\Braket{\mathbf{\hat{F}}(t)}$, which corresponds to the classical description of magnetic spins. Therefore, its dynamics can be written as
\begin{align}
\frac{d\mathbf{P}(t)}{dt}=&\mathbf{B}(t) \mathbf{P}(t)-\Gamma_p(t)\left[\mathbf{P}(t)-\mathbf{P}^{\textrm{in}}\right],\label{eq:dSdtLab}
\end{align}
where we have defined $\mathbf{B}(t)=\mathbf{B}_0(t)+ \mathbf{B}_{\mathrm{ext}}^{(0)}-\Gamma_\mathrm{rel}$
and $\mathbf{P}^{\textrm{in}}=\Braket{\mathbf{\hat{F}}^{\textrm{in}}}$.
The matrix $\mathbf{B}(t)$ can be decomposed spectrally
\begin{align}
\mathbf{B}(t)=\mathbf{B}^{(0)}+&\mathbf{B}^{(1)} e^{i\omega t} +\mathbf{B}^{(-1)}e^{-i\omega t},\label{eq:Blabdecomp}
\end{align}
in which
\begin{align}
\mathbf{B}^{(0)}&=\begin{bmatrix}
    -\Gamma_x & -\Omega_{\mathrm{dc}}-\Omega_z^{\mathrm{ext}} & \Omega_y^{\mathrm{ext}} \\
    \Omega_{\mathrm{dc}}+\Omega_z^{\mathrm{ext}} & -\Gamma_y & -\Omega_x^{\mathrm{ext}} \\
    -\Omega_y^{\mathrm{ext}} & \Omega_x^{\mathrm{ext}} & -\Gamma_z
\end{bmatrix}, \\
\mathbf{B}^{(\pm1)}&=\begin{bmatrix}
    0 & 0 & 0  \\
    0 & 0 & -\frac{\Omega_{\mathrm{rf}}}{2} \\
    0 & \frac{\Omega_{\mathrm{rf}}}{2} & 0
\end{bmatrix}. \label{eq:Bharmonics}
\end{align}
Another term that can be spectrally decomposed is the the pumping rate 
\begin{align}
\Gamma_p(t)=\Gamma_p^{(0)}+& \Gamma_p^{(1)} e^{i\omega t} +\Gamma_p^{(-1)}e^{-i\omega t} \nonumber\\
+& \Gamma_p^{(2)} e^{2i\omega t} +\Gamma_p^{(-2)}e^{-2i\omega t}+\cdots,\label{eq:Ypdecomp}
\end{align}
such that, for instance, a square-wave pumping profile can be decomposed to
\begin{align}
\Gamma_p^{(0)}&=\Gamma_b\ d,\ 
\Gamma_p^{(n)}=\Gamma^{(-n)}=\frac{\Gamma_b}{n\pi}\sin(n\pi d),
\end{align}
where $d$ corresponds to the duty cycle of the carrier wave.
The general description in eq.~(\ref{eq:Ypdecomp}) can simulate a broad range of time dependent pumping rates with different spectral decompositions e.g. sine, sawtooth etc.

With the definitions above, the dynamical equation can be written as
\begin{align}
\frac{d\mathbf{P}(t)}{dt}=&\left(\mathbf{B}^{(0)}+\mathbf{B}^{(1)} e^{i\omega t} +\mathbf{B}^{(-1)}e^{-i\omega t}\right) \mathbf{P}(t)\nonumber\\
&-\sum_n \Gamma_p^{(n)} e^{in\omega t}\ \mathbf{P}(t)+\sum_n \Gamma_p^{(n)} e^{in\omega t}\ \mathbf{P}^{\textrm{in}}.\label{eq:FulldfdtLab2}
\end{align}

To check the consistency of this solution, we show in Appendix \ref{app:RotFrame} that applying the rotating frame transformation and considering no external magnetic fields, we recover the Bloch solution~\cite{Bloch46}. Nevertheless, it is worth noting that eq.~(\ref{eq:FulldfdtLab2}) and its counterpart in the rotating frame in eq.~(\ref{eq:dfdtrot4}) stands for a more general description, which accounts for the presence of external fields and more realistic pumping schemes, which could be harmonically decomposed, as a result, it requires Floquet expansion to solve it.

\subsection{Floquet expansion of the first moment in the laboratory frame \label{sec:Floquet1rst_Lab}}
Given the harmonic nature of this dynamical equation, we employ a Floquet expansion of the spin operators $\mathbf{\hat{F}}(t)$ in 
order to find a steady state solution for all the possible harmonics. Therefore, we expand the spin operator harmonically  as
\begin{align}
\mathbf{\hat{F}}(t)=\mathbf{\hat{F}}^{(0)}(t)+& \mathbf{\hat{F}}^{(1)}(t)\ e^{i\omega t} +\mathbf{\hat{F}}^{(-1)}(t)\ e^{-i\omega t} \nonumber\\
+& \mathbf{\hat{F}}^{(2)}(t)\  e^{2i\omega t} +\mathbf{\hat{F}}^{(-2)}(t)\ e^{-2i\omega t}+\cdots,\label{eq:Flabexpand}
\end{align}
such that for the first moment we have $\mathbf{P}(t)=\sum_n\mathbf{P}^{(n)}(t) e^{in\omega t}$
where $\mathbf{P}^{(n)}(t)=\Braket{\mathbf{\hat{F}}^{(n)}(t)}$.
This expansion allows to compute the dynamics of $\mathbf{P}(t)$
by finding the time evolution of the harmonic components $\mathbf{P}^{(n)}(t)$. Substituting this expansion into eq.~(\ref{eq:FulldfdtLab2}), we find the following dynamics for the spin harmonics

\begin{align}
\frac{d\mathbf{P}^{(n)}}{dt}=\mathbf{B}_n^{(0)}\mathbf{P}^{(n)} &+\mathbf{B}^{(1)}\mathbf{P}^{(n-1)}  + \mathbf{B}^{(-1)}\mathbf{P}^{(n+1)}\nonumber\\
&+\Gamma_p^{(n)} \mathbf{P}^{\textrm{in}}-\sum_{i=-Q}^{Q} \Gamma_p^{(n-i)} \mathbf{P}^{(i)}(t), \label{eq:dSndtLab}
\end{align}
where $\mathbf{B}_n^{(0)}=\mathbf{B}^{(0)}-in\omega \mathbf{I}$ and $Q\in \mathbb{Z}$ corresponds to the cut-off frequency index i.e. a finite, but large number of harmonics required to satisfy convergence in numerical calculations \cite{Levante95,floquet2}. It is worth noting that, in the laboratory frame, the matrices $\mathbf{B}^{(\pm 1)}$ are directly proportional to the amplitude of the rf field, responsible for coupling the harmonics of $\mathbf{P}^{(n)}$ and $\mathbf{P}^{(n\mp1)}$, respectively. However, in Appendix \ref{Sec:Floquete_1stFRot} we show that in the rotating frame, the harmonics are coupled by the matrices $\mathbf{M}^{(\pm 1)}$, which are proportional to the external transverse fields. More specifically, the in-phase component of $\mathbf{M}^{(\pm 1)}$ is proportional to magnetic fields in the $x$-direction, whereas its out of phase component is proportional to the external magnetic field in the $y$-direction. This is going to be described in more detail in the results section.

A compact way to express eq.~(\ref{eq:FulldfdtLab2}) is in a matrix form by defining a new linear space. To do so, we define the harmonic vector $\mathbb{P}=(\cdots,\mathbf{P}^{(-n)}e^{-i n\omega t},\cdots,\mathbf{P}^{(-1)}e^{-i \omega t}$, $\mathbf{P}^{(0)},\mathbf{P}^{(1)}e^{i \omega t},\cdots,\mathbf{P}^{(n)}e^{i n \omega t},\cdots)^T$
 such that, by defining the matrix $\mathbb{N}$ 
 with matrix elements $\mathbb{N}_{ij}=j\delta_{ij}$ and $j$
 spanning for all possible harmonics,
the spin operator can be written as
\begin{align}
\mathbb{P}(t)=e^{i\omega \mathbb{N}t} \mathbb{P}_F, \label{eq:S_hiper}
\end{align}
where $\mathbb{P}_F=(\cdots, \mathbf{P}^{(-n)}, \cdots,\mathbf{P}^{(-1)},\mathbf{P}^{(0)},\mathbf{P}^{(1)},\cdots,\mathbf{P}^{(n)}$, $\cdots)^T$ correspond to the amplitude of the spin harmonic.
Defining the vector $\mathbb{V}$ with $\mathbb{V}_i=1$ for all $i$'s,
the spin operator can be written as $\mathbf{P}(t)=\mathbb{V}\cdot \mathbb{P}(t)$. 
Therefore, the dynamics of the harmonics in eq.~(\ref{eq:dSndtLab}) can be written as 
\begin{align}
\frac{d\mathbb{P}_F}{dt}=[\mathbb{B}-\mathbbm{\Gamma}]\ \mathbb{P}_F +\mathbbm{\Gamma}_{\textrm{in}}\ \mathbb{P}_{\textrm{in}},\label{eq:dSF_hiper_dtLab}
\end{align}
where $\mathbb{B}=\mathbb{B}'-i\mathbb{N}\omega$.
The calculation of the dynamics requires a limitation of this vector by defining a cut-off harmonic $n=Q$ such that the solution converges. In the finite case, we have
\begin{align}
\mathbb{B}'_{nm}=\begin{cases}
      \mathbf{B}^{(0)}-in\omega \mathbf{I}_{3\times3}, & \text{for}\ n=m, \\
      \mathbf{B}^{(\pm1)}, & \text{for}\ m=n\mp1, \\
      0, & \text{otherwise}, \\
    \end{cases}\label{eq:B_hiper}
\end{align}
where the pump matrix elements $(\mathbbm{\Gamma})_{nm}=\Gamma_p^{(n-m)}\mathbf{I}_{3\times3}$, the pump relaxation $(\mathbbm{\Gamma}_{\textrm{in}})_{nm}=\delta_{nm}\Gamma_p^{(n)}\mathbf{I}_{3\times3}$, 
and the input vector $(\mathbb{P}_{\textrm{in}})_n=\mathbf{P}^{\textrm{in}}$.
This new vector has dimensions $d_P=3\times(2Q+1)$, and the matrices have $d_B=d_P\times d_P$. 

From eq.~(\ref{eq:dSF_hiper_dtLab}), the steady state solution is
\begin{align}
\mathbb{P}_F=-[\mathbb{B}-\mathbbm{\Gamma}]^{-1}\mathbbm{\Gamma}_{\textrm{in}}\ \mathbb{P}_{\textrm{in}}. \label{eq:SteadySF_hiper_Lab}
\end{align}
The convergence of this solution is verified when $\mathbb{P}_F$ reaches stability by comparing the calculation with  $\mathbb{B}_{2Q+1}$ and $\mathbb{B}_{2Q+3}$, with $Q$ being the cut-off frequency.

The eq.~(\ref{eq:dSF_hiper_dtLab}) is the general form of the classical solution in eq.~(\ref{eq:dPdtClassical}), where an arbitrary pump intensity profile is considered. In this work, we aim to calculate a more general solution for the spin dynamics, not only showing the first moment solution, but going up to second order moments of the spin dynamics. In particular, we are interested in the solution of the spin co-variance matrix when the atoms are in the presence of an external magnetic field where the spin state is prepared using a synchronous pumping process with a square wave intensity profile. This corresponds to our Voigt effect based 3D vector magnetometer described in ref.~\cite{Tadas19}. 

Before we discuss the first and second moment solutions of the spin operators, it is worth briefly showing the analogy between the dynamics of the first two statistical moments.
We have already shown that in the laboratory frame the first moment follows the dynamics given by
\begin{align}
\frac{d\Braket{\mathbf{\hat{F}}(t)}}{dt}=&\tilde{\mathbf{B}}(t) \Braket{\mathbf{\hat{F}}(t)}
-\Gamma_p(t)\left(\Braket{\mathbf{\hat{F}}(t)}
-\Braket{\mathbf{\hat{F}}^{\textrm{in}}}\right),
\end{align}
where $\tilde{\mathbf{B}}(t)=\mathbf{B}_0+\mathbf{B}_{\textrm{ext}}^{(0)}(t)-\Gamma_\mathrm{rel}$,
whilst the second moment will follow, in the Liouville space, an equivalent dynamics given by
\begin{align}
\frac{d\mathbf{X}(t)}{dt}=&\mathbf{C}(t)\mathbf{X}(t)-2\Gamma_p(t)\ [\mathbf{X}(t)-\mathbf{X}_{\textrm{in}}]+\Lambda_\mathrm{rel}\ \mathbf{X}_{0},
\end{align}
in which $\mathbf{X}(t)$ corresponds to the vector representation of the second moment matrix $\bm{\sigma}(t)=\Braket{\mathbf{\hat{F}}(t)\ \mathbf{\hat{F}}(t)^T}$ in the Liouville space.
This analogy allows us to easily show that the Floquet expansion employed for the first moment solution can be extended to the second moment solution and that the iterative formulas in both cases will be equivalent.

\section{Dynamics of the second moment $\Braket{\mathbf{\hat{F}}(t) \ \mathbf{\hat{F}}(t)^{T}}$ \label{Sec:2ndtMoment}}
\subsection{Dynamics of the second moments in the laboratory frame\label{sec:DynamicsLab}}
Now let us draw our attention to determine the dynamics of the second moment. To do so, we define the second moment matrix
\begin{align}
\bm{\sigma}(t)&=\Braket{\mathbf{\hat{F}}(t)\ \mathbf{\hat{F}}(t)^T}=
\begin{bmatrix}
  \sigma_{xx}(t) & \sigma_{xy}(t) & \sigma_{xz}(t) \\
   \sigma_{yx}(t) & \sigma_{yy}(t) & \sigma_{yz}(t) \\
    \sigma_{zx}(t) & \sigma_{zy}(t) & \sigma_{zz}(t)
\end{bmatrix},\label{eq:sigmalab}
\end{align}
where the matrix elements are $\sigma_{ij}(t)=\Braket{\hat{F}_i(t)\ \hat{F}_j(t)^T}$.
From the spin dynamics we can determine the dynamics of the second moment
\begin{align}
\frac{d\bm{\sigma}(t)}{dt}&=\Braket{\frac{d\mathbf{\hat{F}}(t)}{dt}\ \mathbf{\hat{F}}(t)^T}+\Braket{\mathbf{\hat{F}}(t)\  \frac{d\mathbf{\hat{F}}(t)^T}{dt}},\label{eq:dsigma_dt}
\end{align}
and substituting eq.~(\ref{eq:FulldfdtLabStoch}) and its transpose into eq.~(\ref{eq:dsigma_dt}), we obtain

\begin{align}
\frac{d\bm{\sigma}(t)}{dt}&=\mathbf{B}(t) \bm{\sigma}(t)+\bm{\sigma}(t)\mathbf{B}(t)^T\nonumber\\
&-\Gamma_p(t)\left[2\bm{\sigma}(t)-\Braket{\mathbf{\hat{F}}^{\textrm{in}}} \Braket{\mathbf{\hat{F}}(t)^T}-\Braket{\mathbf{\hat{F}}(t)} \Braket{\mathbf{\hat{F}}^{\textrm{in}~T}}\right]\nonumber\\
&+\left[\Braket{\bm{\mathcal{\hat{F}}}^{\textrm{in}}(t)\ \mathbf{\hat{F}}(t)^T}+\Braket{\mathbf{\hat{F}}(t)\  \bm{\mathcal{\hat{F}}}^{\textrm{in}}(t)^T}\right]\nonumber\\
&+\Braket{\bm{\mathcal{\hat{F}}}(t)\ \mathbf{\hat{F}}(t)^T}+\Braket{\mathbf{\hat{F}}(t)\  \bm{\mathcal{\hat{F}}}(t)^T}.\label{eq:dsigma_dtLab2}
\end{align}

The first two terms on the right hand side of eq.~(\ref{eq:dsigma_dtLab2}) contain the coherent part generated by the interaction between the atomic spin and the magnetic fields. The two terms proportional to the pump rate $\Gamma_p(t)$ describe the dynamics of the second moment due to the pumping process. The last two terms correspond to the contribution of the stochastic noise from the pumping process and unpolarized atoms. 

In particular, we need to determine the cross correlation of the atomic spin $\mathbf{\hat{F}}(t)$ with the input and unpolarized stochastic operators $\bm{\mathcal{\hat{F}}}^{\textrm{in}}(t)$ and $\bm{\mathcal{\hat{F}}}(t)$.
In Appendix \ref{app:cross_correlations}, we show in eq.~(\ref{eq:FFstochLab4}), that the input stochastic term satisfies the following expression
\begin{align}
\Braket{\bm{\mathcal{\hat{F}}}^{\textrm{in}}(t)\ \mathbf{\hat{F}}(t)^T}+\Braket{\mathbf{\hat{F}}(t)\  \bm{\mathcal{\hat{F}}}^{\textrm{in}}(t)^T}=&2\Gamma_p(t) \bm{\sigma}^{\textrm{in}},
\end{align}
and the unpolarized stochastic term is given by eq.~(\ref{eq:FFstochLab6})
\begin{align}
\Braket{\bm{\mathcal{\hat{F}}}(t)\ \mathbf{\hat{F}}(t)^T}+\Braket{\mathbf{\hat{F}}(t)\  \bm{\mathcal{\hat{F}}}(t)^T}=&\mathbf{\Gamma}_\mathrm{rel}\ \bm{\sigma}_{0}+\bm{\sigma}_{0}\ \mathbf{\Gamma}_\mathrm{rel},
\end{align}
such that, substituting the equations above into eq.~(\ref{eq:dsigma_dtLab2}), we finally obtain 
\begin{align}
\frac{d\bm{\sigma}(t)}{dt}=&\mathbf{B}(t) \bm{\sigma}(t)+\bm{\sigma}(t)\mathbf{B}(t)^T
-2\Gamma_p(t)\bm{\sigma}(t)\nonumber\\
&+\mathbf{\Gamma}_\mathrm{rel}\ \bm{\sigma}_{0}+\bm{\sigma}_{0}\ \mathbf{\Gamma}_\mathrm{rel}+2\Gamma_p(t)\ \bm{\sigma}_{\textrm{in}}\nonumber\\
&+\Gamma_p(t)\left[\Braket{\mathbf{\hat{F}}^{\textrm{in}}} \Braket{\mathbf{\hat{F}}(t)^T}+\Braket{\mathbf{\hat{F}}(t)} \Braket{\mathbf{\hat{F}}^{\textrm{in}~T}}\right].\label{eq:dsigma_dtLabPol}
\end{align}
In particular, for polarized samples prepared in an aligned state, $\Braket{\mathbf{\hat{F}}^{\textrm{in}}}=0$, we have 
\begin{align}
\frac{d\bm{\sigma}(t)}{dt}=&\mathbf{B}(t) \bm{\sigma}(t)+\bm{\sigma}(t)\mathbf{B}(t)^T
-2\Gamma_p(t)\left[\bm{\sigma}(t)-\bm{\sigma}_{\textrm{in}}\right]\nonumber\\
&+\mathbf{\Gamma}_\mathrm{rel}\ \bm{\sigma}_{0}+\bm{\sigma}_{0}\ \mathbf{\Gamma}_\mathrm{rel}\label{eq:dsigma_dtLabUnPol}.
\end{align}

\subsection{Second moment matrix dynamics in the Liouville space}
The dynamical equation for the second moment is linearly equivalent to the equation for the first moment. To explicitly show that equivalence we transform
the second moment matrix $\bm{\sigma}(t)$ into a vector $\mathbf{X}(t)$ in the Liouville space such that $\bm{\sigma}(t)\rightarrow \mathbf{X}(t)$.
Meanwhile,  matrices that operate in the Euclidean space from the left and from the right with respect to an operator $\mathbf{\hat{O}}$, i.e. $\mathbf{\hat{O}}\bm{\sigma}_A$ and $\bm{\sigma}_A\mathbf{\hat{O}}$, respectively, are mapped into matrices in the Liouville space such that 
\begin{align}
\mathbf{\hat{O}}\ \bm{\sigma}_A \rightarrow & \mathcal{L}(\mathbf{\hat{O}})\mathbf{X}_A,\\
\bm{\sigma}_A\ \mathbf{\hat{O}} \rightarrow & \mathcal{R}(\mathbf{\hat{O}})\mathbf{X}_A.
\end{align}

Hence, the dynamical eq.~(\ref{eq:dsigma_dtLabUnPol}) can be written in the Liouville space as
\begin{align}
\frac{d\mathbf{X}(t)}{dt}=&\mathbf{C}(t)\mathbf{X}(t)-2\Gamma_p(t)\ [ \mathbf{X}(t)-\mathbf{X}_{\textrm{in}}]
+\Lambda_\mathrm{rel}\ \mathbf{X}_{0},\label{eq:dsigma_dt_LioLab}
\end{align}
where 
\begin{align}
\mathbf{C}(t)&=\mathcal{L}(\mathbf{B}(t))+ \mathcal{R}(\mathbf{B}(t)^T),\label{eq:LiouvilleOperLab}\\
\Lambda_\mathrm{rel}&=\mathcal{L}(\mathbf{\Gamma}_\mathrm{rel})+ \mathcal{R}(\mathbf{\Gamma}_\mathrm{rel}).
\end{align}

From now on we refer to $\mathbf{X}(t)$ as the second moment vector in the Liouville space. We have demonstrated the equivalence between the dynamics of the first and second moment that we pointed out at the end of Sec.~\ref{sec:Floquet1rst_Lab}.

Since the magnetic field interaction matrix $\mathbf{B}(t)$ decomposes
as in eq.~(\ref{eq:Blabdecomp}), in the Liouville space this is equivalently expressed as 
\begin{align}
\mathbf{C}(t)=&\mathbf{C}^{(0)}+ \mathbf{C}^{(1)}\ e^{i\omega t} +\mathbf{C}^{(-1)}\ e^{-i\omega t}, \label{eq:C_expand}
\end{align}
where $\mathbf{C}^{(n)}=\mathcal{L}(\mathbf{B}^{(n)})+\mathcal{R}(\mathbf{B}^{(n)T})$ with $n=-1,0,1$. Again, this can be solved by employing the Floquet expansion.

\section{Floquet expansion of the second moment of the spin operator\label{Sec:Floquet2ndMoment}}

To obtain the Floquet expansion of the dynamical eq.~(\ref{eq:dsigma_dt_LioLab}), we must check how the time dependent variables are harmonically expanded.
Let us start with the harmonic expansion of the second moment matrix, which follows the expansion in eq.~(\ref{eq:Flabexpand}), such that 
\begin{align}
\bm{\sigma}(t)=\bm{\sigma}^{(0)}(t)+& \bm{\sigma}^{(1)}(t)\ e^{i\omega t} +\bm{\sigma}^{(-1)}(t)\ e^{-i\omega t} \nonumber\\
+& \bm{\sigma}^{(2)}(t)e^{2i\omega t} +\bm{\sigma}^{(-2)}(t)\ e^{-2i\omega t}+\cdots,\label{eq:sigam_expandLab}
\end{align}
in which the harmonic component $\bm{\sigma}^{(n)}(t)$ can be generally expressed in terms of the average spin operator products as
\begin{align}
\bm{\sigma}^{(0)}(t)=&\sum_{n=0}[\tilde{\bm{\sigma}}^{(n,-n)}(t)+\tilde{\bm{\sigma}}^{(-n,n)}(t)],\\
\bm{\sigma}^{(q)}(t)=& \sum_{n=0}[\tilde{\bm{\sigma}}^{(n+q,-n)}(t)+\tilde{\bm{\sigma}}^{(-n,n+q)}(t)],\\
\bm{\sigma}^{(-q)}(t)=& \sum_{n=0}[\tilde{\bm{\sigma}}^{(n,-n-q)}(t)+\tilde{\bm{\sigma}}^{(-n-q,n)}(t)],
\end{align}
where $\tilde{\bm{\sigma}}^{(n,m)}(t)=\Braket{\mathbf{\hat{F}}^{(n)}(t)\mathbf{\hat{F}}^{(m)}(t)^T}$.
Therefore, in the Liouville space, where the matrix $\bm{\sigma}(t)$ is 
represented by the vector $\mathbf{X}(t)$, the second moment vector is expanded as
\begin{align}
\mathbf{X}(t)=\mathbf{X}^{(0)}(t)+& \mathbf{X}^{(1)}(t)\ e^{i\omega t} +\mathbf{X}^{(-1)}(t)\ e^{-i\omega t} \nonumber\\
+& \mathbf{X}^{(2)}(t)e^{2i\omega t} +\mathbf{X}^{(-2)}(t)\ e^{-2i\omega t}+\cdots.\label{eq:X_expandLab}
\end{align}

Similarly to the spin polarization $\mathbb{P}$, 
we can now define the vectors in the matrix form for the spectral space 
$\mathbf{X}(t)=\mathbb{V}\cdot\mathbb{X}(t)=\mathbb{V}\cdot e^{i\mathbb{N}\omega t}\ \mathbb{X}_F$
where $(\mathbb{X}_F)_n=\mathbf{X}^{(n)}$.


The Floquet expansion for the dynamical eq.~(\ref{eq:dsigma_dt_LioLab}) is obtain by substituting the spectral expansions of the second moment vector $\mathbf{X}(t)$ and the matrices $\mathbf{C}(t)$
and $\mathbf{\Gamma}_p(t)$. Therefore,  according to eqs.~(\ref{eq:X_expandLab}), (\ref{eq:Ypdecomp}) and  (\ref{eq:C_expand}), and associating terms at the same harmonic frequency, we obtain the following recursive equation
\begin{align}
\frac{d\mathbf{X}^{(n)}(t)}{dt}=&
\mathbf{C}_n^{(0)}\mathbf{X}^{(n)}(t)+\mathbf{C}^{(1)}\mathbf{X}^{(n-1)}(t)\nonumber\\
&+\mathbf{C}^{(-1)}\mathbf{X}^{(n+1)}(t)-2\sum_i \Gamma_p^{(n-i)}\mathbf{X}^{(i)}(t)\nonumber\\
& +  2\Gamma_p^{(n)}\mathbf{X}_{\textrm{in}}+
\Lambda_\mathrm{rel}\ \mathbf{X}_0\ \delta_{n,0}.\label{eq:dXndtLab2}
\end{align}

Notice that the first three terms on the right hand side are equivalent to those obtained for $\mathbf{P}^{(n)}$ for the first order moments.
The spectral convergence of the last term for a given $n^{\mathrm{th}}$-harmonic depends on the decay of the harmonic decomposition of the pumping rate e.g. $\gamma_l^{(n)}\sim \Gamma^{(n)}\sim 1/n$ for a square wave.

As in the case of spins $\mathbf{P}$,
eq.~(\ref{eq:dXndtLab2}) can be expressed in the matrix form as 
\begin{align}
\frac{d\mathbb{X}_F(t)}{dt}=&[\tilde{\mathbb{C}}-\mathbbm{\Gamma}]\ \mathbb{X}_F+ \mathbbm{\Gamma}_{\textrm{in}} \mathbb{X}_{\textrm{in}}+\mathbbm{\Lambda}_\mathrm{rel} \mathbb{X}_{0},
\label{eq:sigam_hiper_expand5}
\end{align}
where $\tilde{\mathbb{C}}=\mathbb{C}-i\mathbb{N}\omega$ takes the same form of $\mathbb{B}$ in eq.~(\ref{eq:B_hiper}), which it components are
\begin{align}
\tilde{\mathbb{C}}_{nm}=\begin{cases}
      \mathbf{C}^{(0)}-in\omega \mathbf{I}, & \text{for}\ n=m, \\
      \mathbf{C}^{(\pm1)}, & \text{for}\ m=n\mp1, \\
      0, & \text{otherwise}, \\
    \end{cases}\label{eq:C_hiper_lab}
\end{align}
where the pump matrix term is $(\mathbbm{\Gamma}_{\textrm{in}})_{nm}=2\delta_{nm}\Gamma_p^{(n)}\mathbf{I}_{9\times9}$, the relaxation matrix $(\mathbbm{\Gamma})_{nm}=\Gamma_p^{(n-m)}\mathbf{I}_{9\times9}$
and with $(\mathbb{X}_{\textrm{in}})_n=\mathbf{X}_{\textrm{in}}$. 
Moreover, the unpolarized drift matrix is $(\mathbbm{\Lambda}_\mathrm{rel})_{nm}=\mathbf{\Lambda}_\mathrm{rel}$ only for 
 $n=m=0$.


Therefore the steady state is governed by 
\begin{align}
 \mathbb{X}_F=&-[\tilde{\mathbb{C}}-\mathbbm{\Gamma}]^{-1}(\mathbbm{\Gamma}_{\textrm{in}} \mathbb{X}_{\textrm{in}}+\mathbbm{\Lambda}_\mathrm{rel}\mathbb{X}_{0}).
\label{eq:SteadySolX}
\end{align}
From this solution we can notice that for a dominant pumping rate, in which $\mathbbm{\Gamma}\gg \tilde{\mathbb{G}},\mathbbm{\Lambda}'_\mathrm{rel}$, the steady state is directly proportional to the input of the second moment vector $\mathbb{X}'_F\approx \mathbbm{\Gamma}^{-1}\mathbbm{\Gamma}'_{\textrm{in}}\mathbb{X}_{\textrm{in}}$, losing any resonant response to the magnetic fields. In the case where collisional processes dominate the dynamics, the solution corresponds to the one for an unpolarized state $\mathbb{X}'_F\approx \tilde{\mathbb{G}}^{-1}\mathbbm{\Lambda}'_\mathrm{rel}\mathbb{X}_{0}$, which also has no resonant response to the magnetic fields.


Furthermore, in the case of the laboratory frame, the first harmonic components $\mathbf{C}^{(\pm1)}$, describe the circular components of the rf field. This does not mean that the steady state solution will contain spin oscillations at the first harmonic when the static field is on resonance. However, approaching the dynamics from the rotating frame we can determine, for some parameters, which element of the dynamics leads to some specific harmonics at resonance. In Sec.~\ref{sec:DynamicsRotFrame} we will discuss the dynamics in the rotating frame. 


\subsection{OPM response during the probe cycle}

\begin{figure}[htb!]
\begin{overpic}[width=8.9cm]{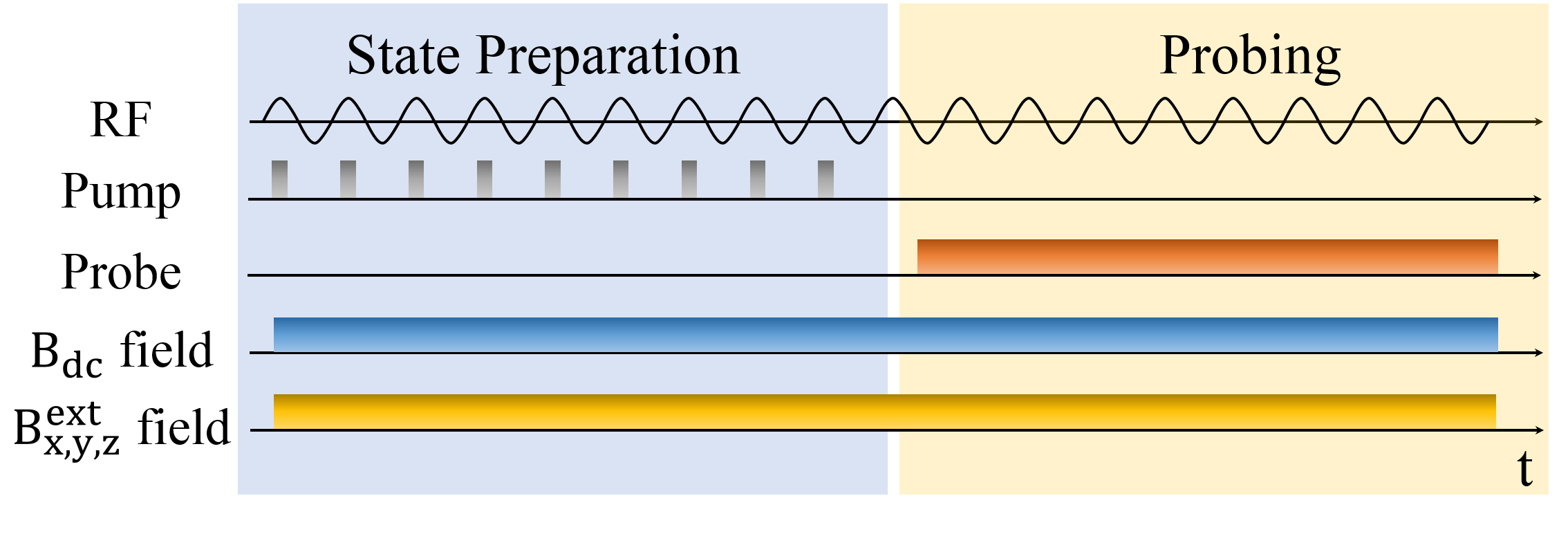}
\end{overpic}
\caption{Free induction decay sequence of Voigt effect magnetometer. The first part of the sequence consists of state preparation (pump cycle) to prepare the aligned states. After the pump cycle a probing pulse interrogates the free induction decay dynamics. This configuration is defined as a double-step measurement. On the other hand, probing during the state preparation is defined as a single-step measurement.}
\label{fig:Pulse_sequence}
\end{figure}

From the steady state in eq.~(\ref{eq:SteadySolX}) one can notice that the pump rate matrix broadens the magnetic response by the Lorentzian term $[\tilde{\mathbb{C}}-\mathbbm{\Gamma}]^{-1}$. To avoid broadening, we utilise pump-probe strategy as it is shown in Fig.~\ref{fig:Pulse_sequence}, where the pumping is followed by a free induction decay of the spin evolution, which is where the state is probed.

Therefore, after a pumping cycle, we have a cycle of atomic dynamics without pumping, such that
\begin{align}
\frac{d\mathbb{X}_F(t)}{dt}=&\tilde{\mathbb{C}}\ \mathbb{X}_F+ \mathbbm{\Lambda}_\mathrm{rel} \mathbb{X}_{0},
\label{eq:sigam_hiper_expand6}
\end{align}
integrating this expression yields the time evolution
\begin{align}
\mathbb{X}_F(t)=&e^{\tilde{\mathbb{C}}\ t}\ \mathbb{X}_F(0)+ \tilde{\mathbb{C}}^{-1}(e^{\tilde{\mathbb{C}}\ t}-\mathbb{I})\ \mathbbm{\Lambda}_\mathrm{rel} \mathbb{X}_{0}.
\label{eq:SolXevol}
\end{align}

To determine the amplitude of the harmonics at the beginning of the probe cycle we integrate over a cycle of the rf frequency, such that the real and imaginary parts for a given harmonic $n$ are
\begin{align}
\mathbf{X}_R^{(n)}&=\frac{1}{T}\int_0^T\cos(n\omega t')\mathbf{X}(t')dt',\\
\mathbf{X}_I^{(n)}&=\frac{1}{T}\int_0^T\sin(n\omega t')\mathbf{X}(t')dt',
\end{align}
where $T=2\pi/\omega$. Substituting the expansion $\mathbf{X}(t)=\sum_{n=-Q}^Q e^{i\omega n t} (\mathbb{X}_F(t))_n$
for the real part we have
\begin{align}
\mathbf{X}_{R}^{(n)}=\frac{1}{T}\int_0^T dt'\sum_{m=-Q}^Q \cos(n\omega t')e^{i\omega m t'} (\mathbb{X}_F(t'))_m.\label{eq:XR}
\end{align}

Since $(\mathbb{X}_F(t'))_m$ are slow varying envelopes of the harmonics within a time $T$, they are even, and therefore, the only non-zero values are
\begin{align}
\mathbf{X}_{R}^{(n)}=\frac{1}{2T}\int_0^T dt' \left[(\mathbb{X}_F(t'))_{n}+(\mathbb{X}_F(t'))_{-n}\right].\label{eq:XR2}
\end{align}

Following the same procedure, the imaginary part is
\begin{align}
\mathbf{X}_I^{(n)}&=-\frac{i}{2T}\int_0^T[(\mathbb{X}_F(t'))_{n}-(\mathbb{X}_F(t'))_{-n}]dt'.\label{eq:XI2}
\end{align}

As a result, we need the integration of the elements $\mathbf{X}^{(n)}(t')$. From the solution in eq.~(\ref{eq:SolXevol}) and  considering that the initial second moment matrix is given by the steady state solution $\mathbb{X}_F(0)=\mathbb{X}_F^s$ in eq.~(\ref{eq:SteadySolX}),
we obtain
\begin{align}
\frac{1}{T}\int_0^T dt'\mathbb{X}_F(t)&=\frac{1}{T}
(\tilde{\mathbb{C}})^{-1} \left(e^{\tilde{\mathbb{C}}\ T} - \mathbbm{1}\right) \mathbb{X}_F^s\nonumber\\
&+\frac{1}{T}
(\tilde{\mathbb{C}})^{-1}\left[(\tilde{\mathbb{C}})^{-1}\left(e^{\tilde{\mathbb{C}}\ T} - \mathbbm{1}\right)-T\right]\mathbbm{\Lambda}_\mathrm{rel}\mathbf{X}_0.\label{eq:XlabtDem}
\end{align}
From this solution one can determine the specific harmonics that contribute to the change in ellipticity in eq.~(\ref{eq:Voigt}). Since the Floquet expansion is applied to the the atomic term $\langle \hat{F}_x^2-\hat{F}_y^2\rangle$,  the ellipticity in eq.~(\ref{eq:Voigt}) can be equally expanded as

\begin{align}
\Braket{\hat{S}_z'(t)}=
G_{F}^{(2)} S_y n_F \sum_{n=-\infty}^{\infty}&\left(h_X^{(n)}+i~h_Y^{(n)}\right)e^{in\omega t}\label{eq:Voigt_floquet},
\end{align}
where the harmonic components $h_{X,Y}^{(n)}$ correspond to the quadratures that can be measured using lock-in detection and defined in terms of eqs.(\ref{eq:XR2}) and (\ref{eq:XI2}) as
\begin{align}
h_X^{(n)}&=[\mathbf{X}_{R}^{(n)}]_1-[\mathbf{X}_{R}^{(n)}]_5\\
h_Y^{(n)}&=[\mathbf{X}_{I}^{(n)}]_1-[\mathbf{X}_{I}^{(n)}]_5\label{eq:h_n},
\end{align}
in which $[\mathbf{X}_{R}^{(n)}]_1$ and $[\mathbf{X}_{I}^{(n)}]_1$
correspond to the real and imaginary components of the nth-harmonic and  the subscript $1$ represents  $\langle \hat{F}_x^2\rangle$, whereas the subscript $5$ represents  $\langle \hat{F}_y^2\rangle$. In particular, 
we are interested in the quadratures of the first harmonics $h_X^{(1)}$ and $h_Y^{(1)}$,  and the real part of the second harmonic $h_X^{(2)}$, because those are the signals that can map the three components of the external magnetic field \cite{Tadas19}.

Notice that the the steady state solution in eq.~(\ref{eq:SteadySolX}) and the dynamical solution in eq.~(\ref{eq:SolXevol}), with the use of the harmonic space can be used to construct a broad range of pumping and probing schemes (single step or double step measurement), as well as include various forms of external magnetic field interactions.
This shows how the algebraic complexity of the coupling of the harmonics seen from the Liouville space, is effectively cleared in the harmonic space, allowing us to distinguish the contribution of different elements in the atomic spin dynamics.



\subsection{Second moment dynamics in the rotating Frame\label{sec:DynamicsRotFrame}}
Although all the numerical results that are going to be presented in the following sections are calculated in the laboratory frame (due to its easier computational implementation), it is nevertheless worth exploring the analytical solution in the rotating frame in order to develop a physical insight into the magnetometer response.

The main feature of this vector-OPM is that the mode quadratures of the first harmonic map the transverse fields. 
In order to show this, let us consider the transformation 
in eq.~(\ref{eq:dfdtRotframeApp}) applied to the second moment matrix in the laboratory frame in eq.~(\ref{eq:sigmalab})
\begin{align}
\bm{\sigma}'(t) &=\mathbf{R}(t)^{-1}\ \bm{\sigma}(t)\  \mathbf{R}(t)\label{eq:sigmarot}.
\end{align}

The corresponding dynamics of the second moment matrix in the rotating frame is described in Appendix \ref{app:sigma_rot_transform}. In the Liouville space the dynamics of the second moment are given by eq.~(\ref{app:dsigma_dt_LioRot}) 


\begin{align}
\frac{d\mathbf{X}'(t)}{dt}=&\mathbf{G}(t)\mathbf{X}'(t)-2[\Gamma_p(t)~\mathbf{X}'(t)-\mathbf{\Gamma}'_p(t)\mathbf{X}_{\textrm{in}}]
+\Lambda'_\mathrm{rel}\ \mathbf{X}_{0},\label{eq:dsigma_dt_LioRot}
\end{align}
where $\mathbf{G}(t)$ describes the resonant and external fields, 
$\mathbf{\Lambda}'_\mathrm{rel}$ accounts for collisional relaxation rates and pump relaxation rate $\mathbf{\Gamma}'_p(t)$ in the rotating frame defined in eq.~(\ref{eq:gammaRot_n}) 


Similarly to what was shown in Sec.~\ref{Sec:2ndtMoment} and \ref{Sec:Floquet2ndMoment},
the $\mathbf{G}(t)$ decomposes as 
$\mathbf{G}(t)=\mathbf{G}^{(0)}+ \mathbf{G}^{(1)}\ e^{i\omega t} +\mathbf{G}^{(-1)}\ e^{-i\omega t}$, 
where $\mathbf{G}^{(n)}=\mathcal{L}(\mathbf{M}^{(n)})+\mathcal{R}(\mathbf{M}^{(n)T})$ with $n=-1,0,1$, and in particular, $\mathbf{G}^{(\pm1)}$ is only dependent on the transverse external fields according to the definition of $\mathbf{M}^{(\pm1)}$ in eq.~(\ref{eq:M0pm1}).
Therefore, the Floquet expansion for the dynamics of $\mathbf{X}'(t)$ is given by
\begin{align}
\frac{d\mathbb{X}'_F(t)}{dt}=&[\tilde{\mathbb{G}}-\mathbbm{\Gamma}]\ \mathbb{X}'_F+ \mathbbm{\Gamma}'_{\textrm{in}} \mathbb{X}_{\textrm{in}}+\mathbbm{\Lambda}'_\mathrm{rel} \mathbb{X}_{0},
\label{eq:sigamrot_hiper_expand}
\end{align}
where
\begin{align}
\tilde{\mathbb{G}}_{nm}=\begin{cases}
      \tilde{\mathbf{G}}^{(0)}-in\omega \mathbf{I}, & \text{for}\ n=m, \\
      \mathbf{G}^{(\pm1)}, & \text{for}\ m=n\mp1, \\
      0, & \text{otherwise}, \\
    \end{cases}\label{eq:C_hiper_lab}
\end{align}
whilst  the pump matrix term is $(\mathbbm{\Gamma}'_{\textrm{in}})_{nm}=2\delta_{nm}
{\mathbf{\Gamma}^\prime_p}^{(n)}$,
the relaxation matrix $(\mathbbm{\Gamma})_{nm}=2\Gamma_p^{(n-m)}\mathbf{I}_{9\times9}$
and the input second moment vector $(\mathbb{X}_{\textrm{in}})_n=\mathbf{X}_{\textrm{in}}$. 
The pump elements ${\mathbf{\Gamma}'_p}^{(n)}$ are defined in Appendix E, eq.~(\ref{eq:gammaRot_n}).
Moreover, the unpolarized drift matrix is
$(\mathbbm{\Lambda}'_\mathrm{rel})_{nm}= \mathbf{\Lambda}'_\mathrm{rel}$ for $ n=m=0$ 
otherwise is zero. Hence, the steady state solution in the rotating frame is
\begin{align}
\mathbb{X}'_F=&[\tilde{\mathbb{G}}-\mathbbm{\Gamma}]^{-1}[ \mathbbm{\Gamma}'_{\textrm{in}} \mathbb{X}_{\textrm{in}}+\mathbbm{\Lambda}'_\mathrm{rel} \mathbb{X}_{0}],
\label{eq:sigamrot_hiper_steadysol}
\end{align}
which takes the same form as the steady state in the laboratory frame.

Transforming back to the laboratory frame (see Appendix ~\ref{app:RotTrans_sigma}), from eq.~(\ref{eq:RLab2}), the solution is
$ \mathbf{X}(t) =\sum_n\mathbf{X}^{(n)}(t) e^{in\omega t}$
with
 \begin{align}
\mathbf{X}^{(n)}(t) &=
\mathsf{R}^{(0)}\mathbf{X}'^{(n)}(t)+\mathsf{R}^{(1)}\mathbf{X}'^{(n-1)}(t)\nonumber\\
&\hspace{3cm}+\mathsf{R}^{(-1)}\mathbf{X}'^{(n+1)}(t)\nonumber\\
&+\mathsf{R}^{(2)} \mathbf{X}'^{(n-2)}(t) +\mathsf{R}^{(-2)}\mathbf{X}'^{(n+2)}(t).\label{eq:XLab_n2}
\end{align}
where $\mathsf{R}^{(n)}$ are Liouville harmonic rotation matrices.

For a general case it is not straight forward to specify the contribution of the processes into the detection of atomic response at a particular harmonic frequency. Nevertheless, we can discuss different scenarios where we can clearly determine the appearance of harmonics.

First, notice that in the rotating frame, the matrices $\mathbf{G}^{(\pm 1)}$ couple the harmonics among them, which are directly mapped onto the $\mathbf{M}^{(\pm1)}$ matrices. According to eq.~(\ref{eq:M0pm1}), these matrices depend directly on the transverse fields, in contrast to the lab frame, in which the $\mathbf{M}^{(\pm1)}$ matrices are only dependent on the rf amplitude. The fact of $\mathbf{M}^{(\pm1)}$ matrices map directly the presence of the transverse fields is consistent with the experimental observation, since only transverse field can give rise to the first harmonic.

So, let us consider the case where there are no transverse fields and we have a constant pump in the system i.e. $\mathbf{G}^{(\pm 1)}=0$ and $\Gamma_p^{(n)}=\Gamma_P\delta_{n,0}$. This condition decouples the harmonics in eq.~(\ref{eq:sigamrot_hiper_expand}), diagonalizes the matrix $\tilde{\mathbb{G}}$ and therefore, the steady state solution is 
\begin{align}
\mathbf{X}'^{(0)}&=[\tilde{\mathbf{G}}^{(0)}-\Gamma_p^{(0)}\mathbf{I}]^{-1}[\tilde{\mathbf{\Gamma}}_p^{(0)} \mathbf{X}_{\textrm{in}}+\mathbf{\Lambda}'_\mathrm{rel} \mathbf{X}_{0}],
\label{eq:sigamrot_0_steadysol}\\
\mathbf{X}'^{(n)}&=0,\ \text{for } n\neq0, 
\end{align}
for the aligned input state along the rf-axis.
Since the only non-zero solution is $\mathbf{X}'^{(0)}$, according to eq.~(\ref{eq:XLab_n2}),
the only harmonics present in the system are
\begin{align}
\mathbf{X}^{(0)}&=\mathsf{R}^{(0)}\mathbf{X}'^{(0)},\label{eq:XLab_n3}\\
\mathbf{X}^{(\pm1)}&=\mathsf{R}^{(\pm1)}\mathbf{X}'^{(0)},\label{eq:XLab_n4}\\
\mathbf{X}^{(\pm2)} &=\mathsf{R}^{(\pm2)}\mathbf{X}'^{(0)}.\label{eq:XLab_n5}
\end{align}

It can be shown that the transverse components of $\mathbf{X}^{(\pm1)}$ are zero due to matrices $\mathsf{R}^{(\pm1)}$.
Therefore, the birefringence from the atoms is described only by $\mathbf{X}^{(0)}$ and $\mathbf{X}^{(\pm2)}$. However, when weak transverse fields are present such that $\mathbf{G}^{(\pm 1)}$ take non-zero values, the next contribution to the solution in eq.~(\ref{eq:XLab_n2}) are $\mathbf{X}'^{(\pm1)}$ and then 
\begin{align}
\mathbf{X}^{(1)}\approx \mathsf{R}^{(0)}\mathbf{X}'^{(\pm1)},\label{eq:XLab_Bxy}
\end{align}
which represents the presence of the weak signal at $\omega$. This is how, for low external fields, the quadratures at $\omega$ directly map the external transverse fields.


\begin{figure*}[t!]
\begin{overpic}[width=\textwidth]{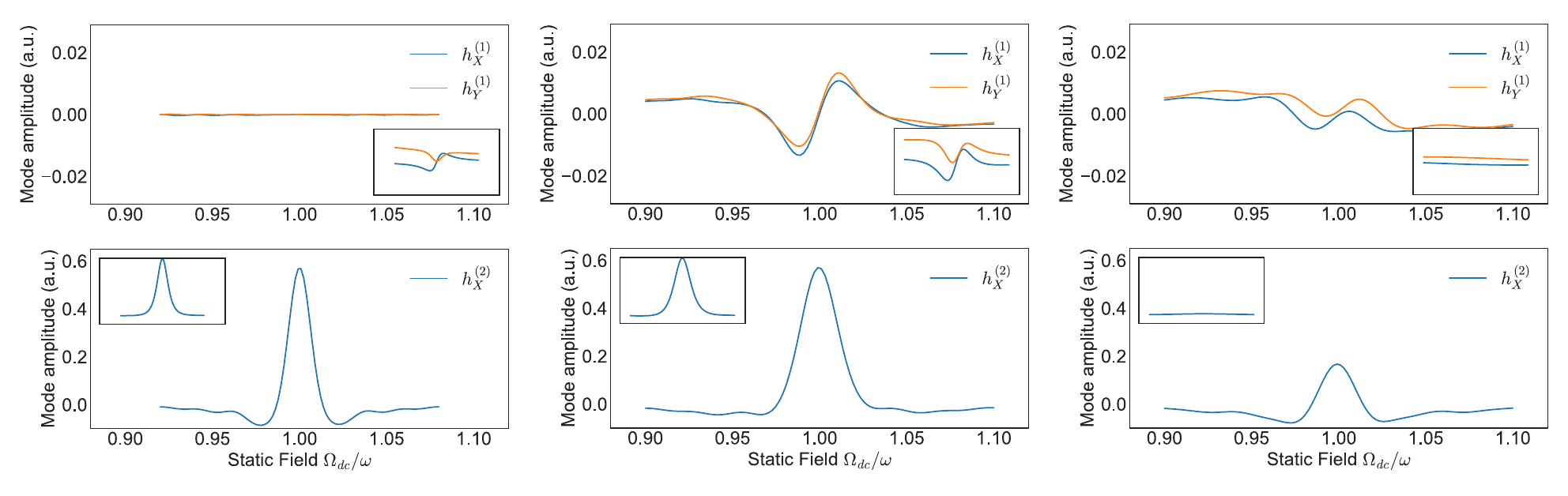}
\put (0,32) {(a)} \put (34,32) {(b)} \put (70,32) {(c)} 
\end{overpic}
\caption{Theoretical behaviour of the harmonics $h_X^{(1)}$, $h_Y^{(1)}$ and $h_X^{(2)}$ as a function of the normalized static $B_z$ field. 
Here (a) and (b) correspond to the situation with no external field $\Omega_\mathrm{ext}=0$ and with $\Omega_x^\mathrm{ext}=\Omega_y^\mathrm{ext}=0.03~\omega_\textrm{rf}$, considering a synchronous pump with a square modulated intensity profile with a 10\% duty cycle at $\omega_\textrm{rf}/2\pi=5$kHz rf-dressing frequency. 
The small insets show a more detailed structure of the $h_x$, $h_y$ and $h_z$ quadratures as well as pump pulse sequences of state preparation and probing. Additional parameters used in the calculation: Input covariance matrix for the aligned state $\rho=(1/2)(|F=2,m_F=2\rangle_x\langle F=2,m_F=2|_x+|F=2,m_F=-2\rangle_x\langle F=2,m_F=-2|_x)$, $\Gamma_b=0.01~\omega_\textrm{rf}$, $\Gamma_i=0.001~\omega_\textrm{rf}$ with $i=x,y,z$, $\Omega_x^\mathrm{ext}=\Omega_y^\mathrm{ext}=0.03~\omega_\textrm{rf}$ and cut-off frequency index Q=5. }
\label{fig:profiles}
\end{figure*}


\section{Results \label{Sec:Results}}
Here we aim to analyze the second moments of state in the hyperfine level $F=2$ of alkali atoms like Rubidium which we can relate to experimental observations in our previous work~\cite{Tadas19}.For this hyperfine level the cut-off frequency index is $Q=5=2F+1$ for which the computation converges. Nevertheless, the model can be applied to other atomic species with different hyperfine level structure.

\subsection{Second moments for aligned states}\label{sec:SecondMom_Aligned}


Let us first analyze the dynamics involving synchronous pumping, in particular, for a stretched state along the $x$-direction, which corresponds to a mixture of equally populated states $|F=2,m_F=\pm2\rangle_x$   with its second moment vector given by
$\mathbf{X}_\textrm{in}=(4,0,0,0,1,0,0,0,1)^T$.
Figure~\ref{fig:profiles} shows the harmonics $h_{X,Y}^{(1)}$ and $h_{X}^{(2)}$ as a function of the static field with its Larmor frequency $\Omega_\textrm{dc}$.
The curves describe the second moment dynamics  after the state preparation in a double-step measurement as it is shown in Fig.~\ref{fig:Pulse_sequence}. The insets show the dynamics of the observables probed in the steady state during the state preparation, which corresponds to a simultaneous pump and probe measurement (single-step measurement). 
Figure~\ref{fig:profiles} compares the results from eq.~(\ref{eq:XlabtDem}) for three different situations: (a) and (b) with synchronous pumping at $d=10\%$ duty cycle with and without the transverse external fields; (c) with continuous-wave (cw) and transverse external fields.



\begin{figure}[b!]
\begin{overpic}[width=8.6cm]{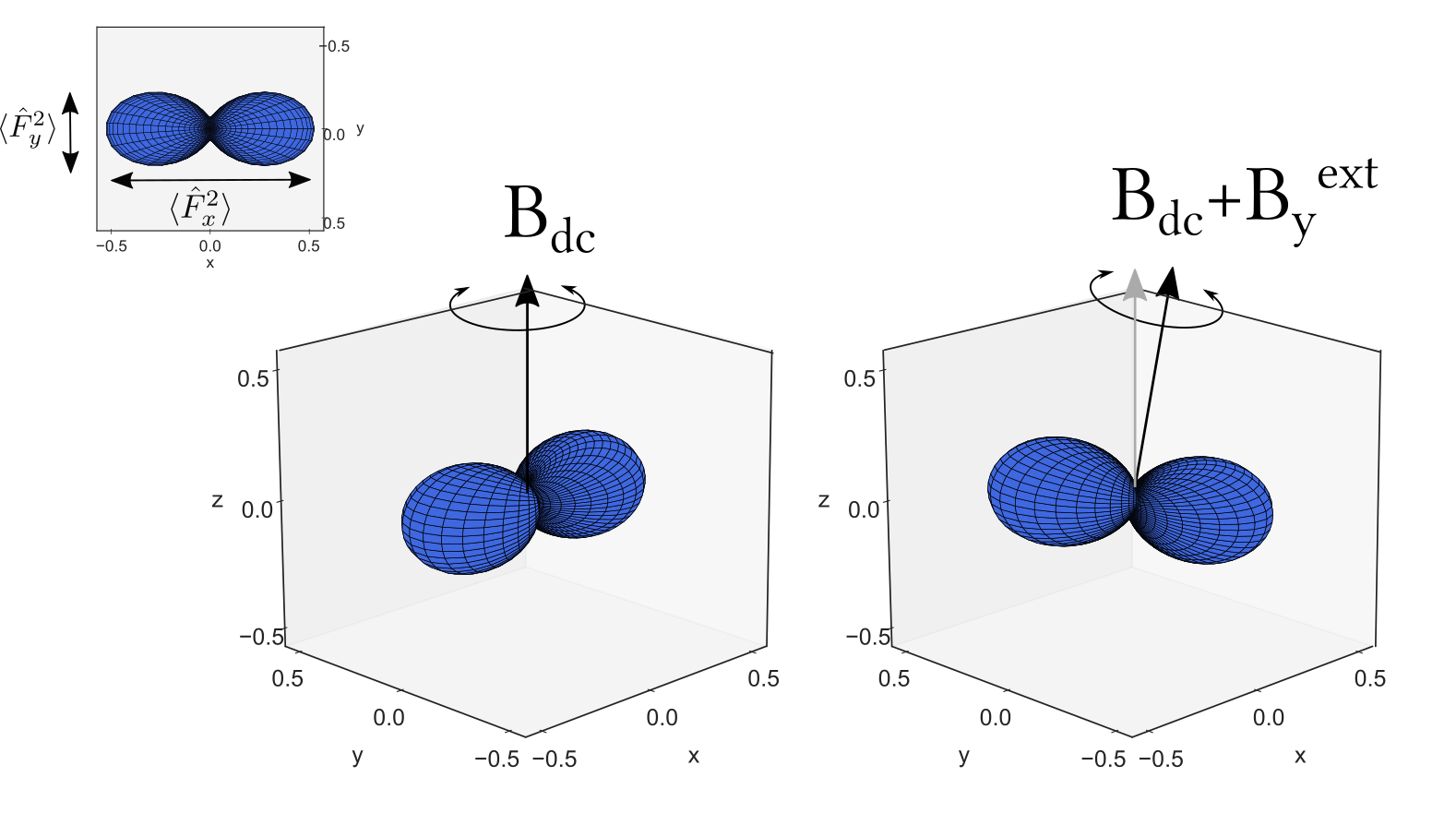}
\put (0,30) {(a)}\put (60,50) {(b)}
\end{overpic}
\caption{(a) Probability surface representation~\cite{polarised_atoms} of the stretched state dynamics (a) without external transverse fields and (b) with external fields tilting the axis of rotation. Inset: top view of the stretched state along the $x$-axis showing that $\langle \hat{F}_x^2-\hat{F}_y^2\rangle>0$.}
\label{fig:ProbSurf_stretched}
\end{figure}

In the case of no external fields in Fig.~\ref{fig:profiles}~(a), one can notice that   the real and imaginary part of the first harmonic are zero, whereas the  real part of the second harmonic $h_X^{(2)}$ has a resonant profile, reaching its maximum when the Larmor frequency $\Omega_\textrm{dc}=\omega_\textrm{rf}$.
This is consistent with the results in sec.~\ref{sec:DynamicsRotFrame}, eq.~(\ref{eq:sigamrot_0_steadysol}) where the absence of transverse fields gives rise to rf signals only in the second harmonic. This can be described in terms of the state probability surface as shown in Fig.~\ref{fig:ProbSurf_stretched}~(a). In this case, the stretched state precesses around the static field which is perfectly aligned with the quantization axis. After half a cycle, the surface returns to its initial position, in which $\langle \hat{F}_x^2\rangle-\langle \hat{F}_y^2\rangle>0$, hence oscillating at twice the Larmor frequency.

In the case of a pump beam with a time dependent square intensity profile which is synchronously modulated as shown in Fig.~\ref{fig:profiles}~(b), with the static field being resonant with the rf driving frequency, the quadratures of the first harmonic follow a dispersive profile due to the presence of weak transverse fields, whereas the second harmonic profile shows resonant behavior, 
describing very closely what was observed experimentally \cite{Tadas19}.
From the rotating frame description, eq.~(\ref{eq:XLab_Bxy}) shows that for weak transverse fields the first harmonic becomes non-zero. Fig.~\ref{fig:ProbSurf_stretched}~(b) shows that in this case the weak transverse fields slightly tilts the static fields axis inducing a precession of the stretched state. In this situation the second harmonic dominates, since the projection $\langle \hat{F}_x^2\rangle-\langle \hat{F}_y^2\rangle>0$  oscillates at twice the Larmor frequency, but the axis of the aligned state only returns to its initial position after an entire Larmor cycle.

In contrast to the modulated pump, in the case of cw pumping, we have $\Gamma^{(0)}\neq0$ and $\Gamma^{(n)}=0$ for $n> 0$. 
Fig.~\ref{fig:profiles}~(c) shows that the first harmonic quadratures no longer display dispersive characteristics and the amplitude of the second harmonic is reduced. This means that spin evolution is not sensitive to the magnetic fields in all three directions.



\begin{figure}[t!]
\begin{overpic}[width=8.6cm]{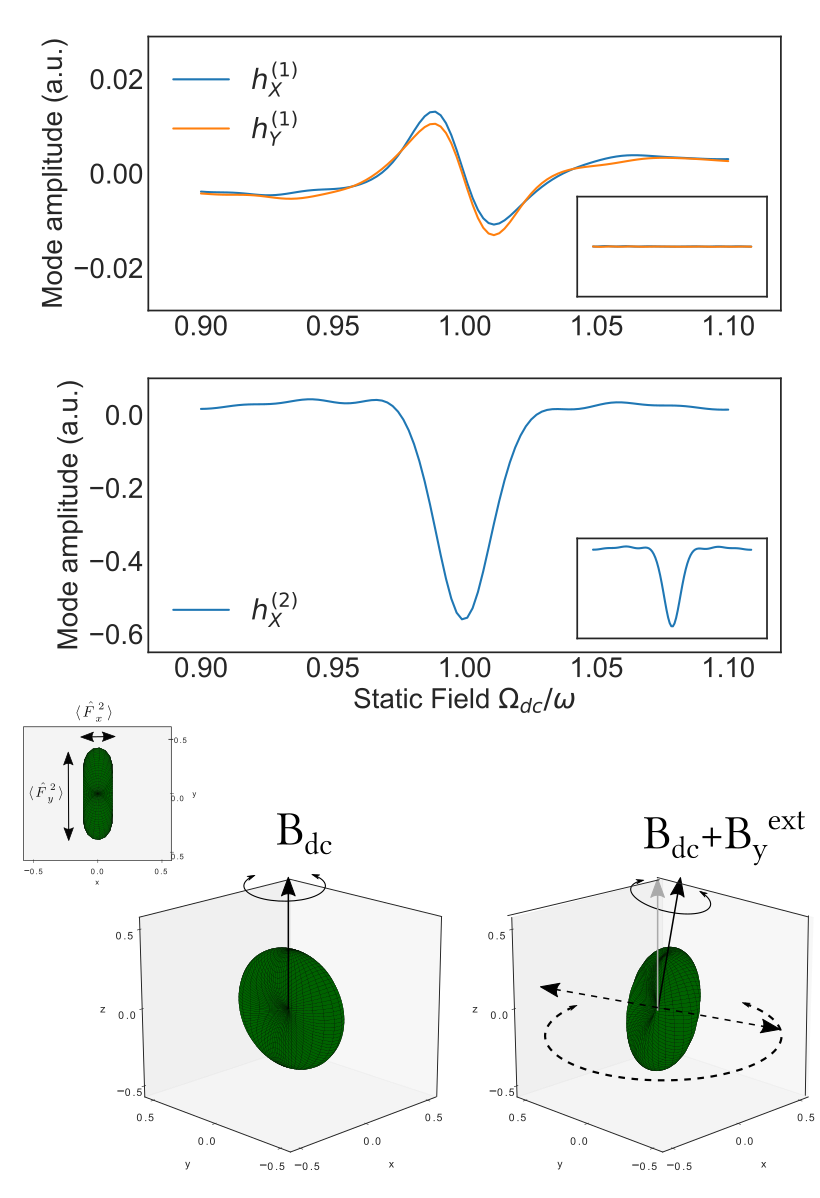}
\put (-2,95) {(a)}\put (-2,68) {(b)}\put (0,20) {(c)}\put (40,28) {(d)}
\end{overpic}
\caption{(a) Theoretical behaviour of the harmonics $h_X^{(1)}$, $h_Y^{(1)}$ and $h_X^{(2)}$ as a function of the normalized static $B_z$ field for an aligned state $|F=2,m_F=0\rangle_x$ . Here (a) and (b) correspond to the situation with external field $\Omega_x^\mathrm{ext}=\Omega_y^\mathrm{ext}=0.03~\omega_\textrm{rf}$, considering the same parameters as in Fig.~\ref{fig:profiles}. (c) and (d) Probability surfaces without and with external transverse fields. Inset: top view of the stretched state along the $x$-axis showing that $\langle \hat{F}_x^2-\hat{F}_y^2\rangle<0$.}
\label{fig:disk_1f_2f_profile}
\end{figure}
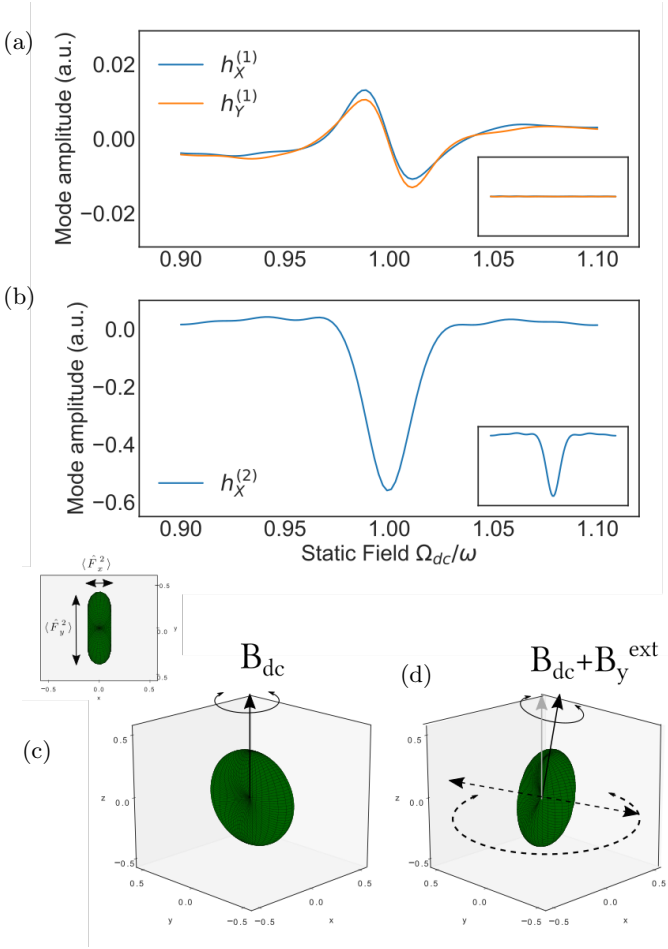

Now we turn to the analysis a second example. Another type of an aligned state is a pure state given by $|F=2,m_F=0\rangle_x$ (see Fig.~\ref{fig:disk_1f_2f_profile}~(c)) where its second moment vector is $\mathbb{X}_{in}=(0,0,0,0,3,0,0,0,3)^T$.
Fig.~\ref{fig:disk_1f_2f_profile}~(a) and (b) show the mode profiles of the first and second harmonics, with the presence of external fields in both transverse directions.
As in the stretched state case, this aligned state has a non-zero second moment difference. The dispersive and  resonant profiles are very similar to those of the stretched states in Fig.~\ref{fig:profiles}~(b), except with an opposite sign. 
The inset shows the mode amplitude for no external magnetic field, in which the only non-zero rf signal is the second harmonic. Similar to the stretched state in Fig.~\ref{fig:ProbSurf_stretched}, the aligned state $|2,0\rangle_x$  precesses around the static field maximizing the $\langle \hat{F}_x^2\rangle-\langle \hat{F}_y^2\rangle$ which oscillates only at twice the Larmor frequency (see Fig.~\ref{fig:disk_1f_2f_profile}~(c)). Once the external transverse field is present, the axis of symmetry is tilted, and its axis precession contributes to rf-signal at the Larmor frequency. It is worth noting that the in-phase top view of the aligned state (insets) is 90 degrees rotated with respect to the stretched state in Fig~\ref{fig:disk_1f_2f_profile}~(c), which shows that $\langle \hat{F}_x^2\rangle-\langle \hat{F}_y^2\rangle<0$ and therefore the mode amplitudes $h_{X,Y}^{(n)}$ have the opposite sign.

\subsection{Second moments for oriented states}
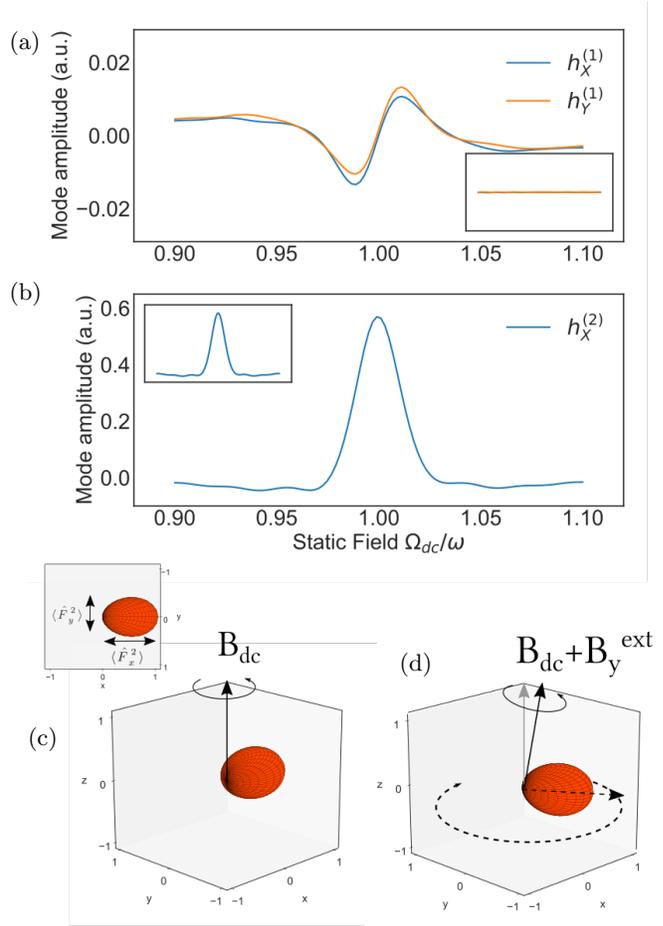
\begin{figure}[b!]
\begin{overpic}[width=8.6cm]{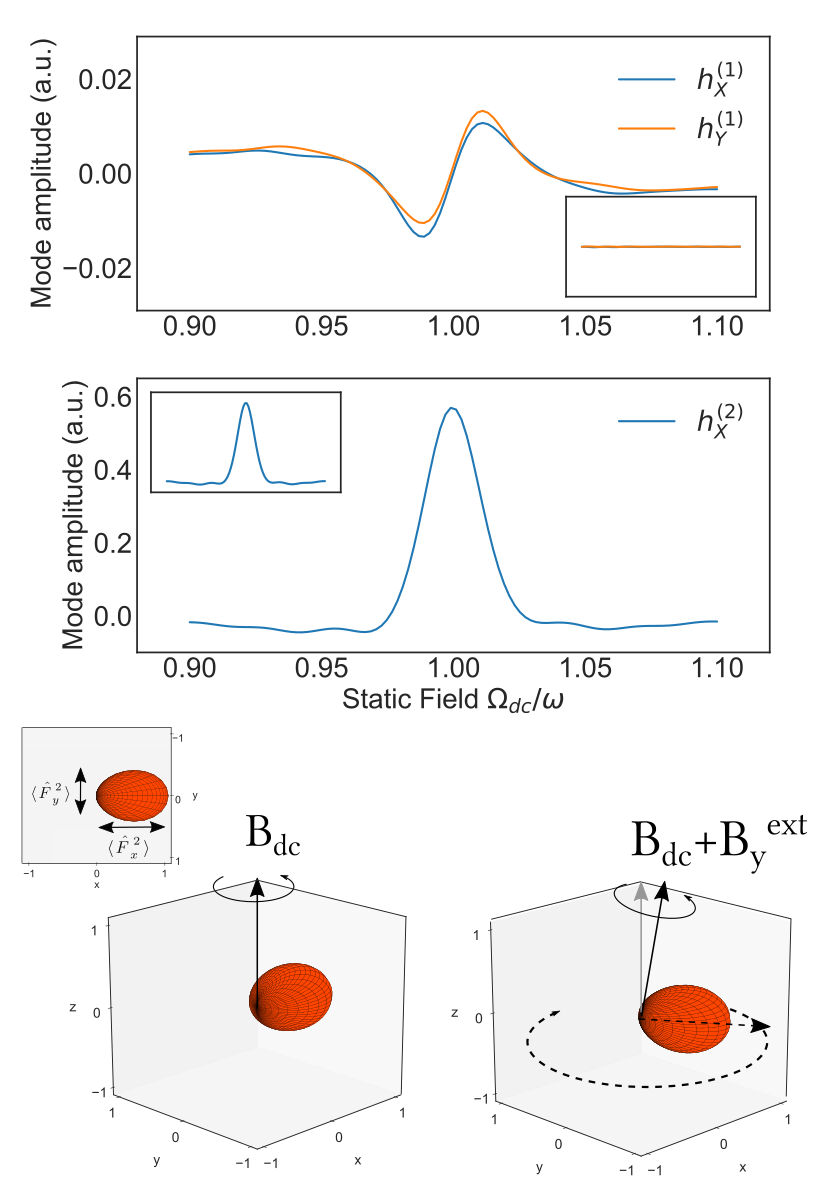}
\put (-2,95) {(a)}\put (-2,68) {(b)}\put (0,20) {(c)}\put (40,28) {(d)}
\end{overpic}
\caption{Theoretical behaviour of the harmonics $h_X^{(1)}$, $h_Y^{(1)}$ and $h_X^{(2)}$ as a function of the normalized static $B_z$ field for an oriented state $|F=2,m_F=2\rangle_x$. Here (a) and (b) correspond to the situation with  external field  $\Omega_x^\mathrm{ext}=\Omega_y^\mathrm{ext}=0.03~\omega_\textrm{rf}$, considering same parameters as in Fig.~\ref{fig:profiles}.}
\label{fig:cone_1f_2f_profile}
\end{figure}

Another effective state that maximizes the precession of the second moments is a transverse oriented state. In particular, consider $|F=2,m_F=2\rangle_x$  where its second moment vector is given by $\mathbb{X}_{in}=(4,0,0,0,1,-i,0,i,1)^T$, very similar to the one for the stretched state. Fig.~\ref{fig:cone_1f_2f_profile} shows the first and second harmonic as a function of the static field for an oriented state along the $x$-axis synchronously pumped, which achieves the same oscillation amplitude as an aligned state. The dispersive profile of $h_{X,Y}^{(1)}$ is non-zero due to transverse fields, whereas the inset shows zero first harmonic response with no transverse field, as in the case of both aligned states discussed above. Fig.~\ref{fig:cone_1f_2f_profile}~(c) and (d) show the representation of the oriented state precessing with and without a transverse field, in the same manner as the aligned state. The in-phase top view of the aligned state (inset) is not rotated with respect to the stretched state in Fig.~\ref{fig:ProbSurf_stretched}, which shows that $\langle \hat{F}_x^2\rangle-\langle \hat{F}_y^2\rangle>0$, and the signs of the signal are not inverted with respect to the stretched state.

The examples considered here with these three states show the atomic spin interaction with the rf field and the external field. One thing to notice is that for optimum sensitivity the aligned states are generated with a parallel pump-probe configuration, whilst the oriented stated requires an orthogonal pump-probe configuration for its optimal preparation. Hence, the aligned state are more convenient for developing miniature sensors, since they can be operated in parallel single-axis configuration.

\subsection{Optimization}
\begin{figure}[t!]
\begin{overpic}[width=8.6cm]{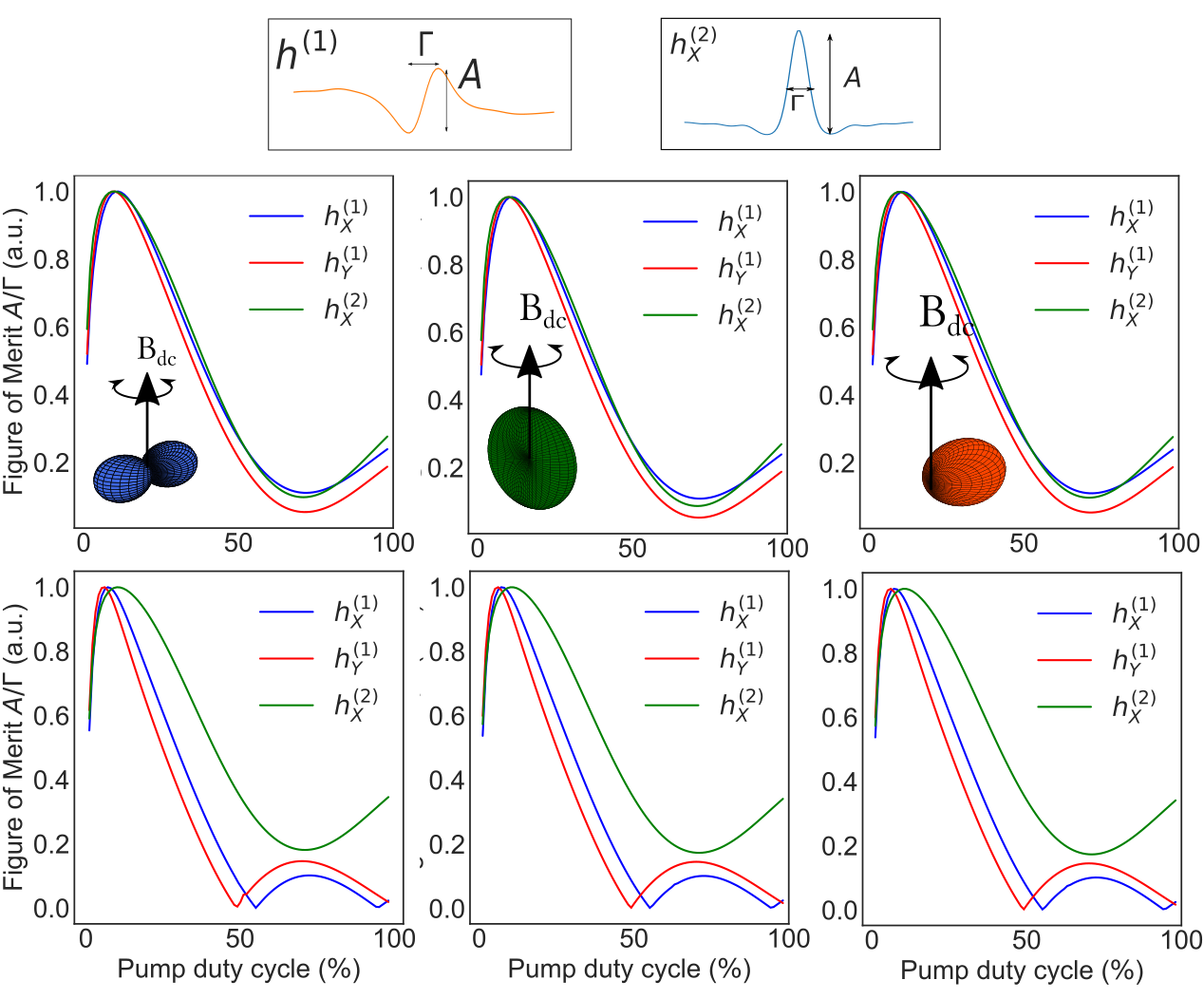}
\put (-3,70) {(a)}\put (-3,35) {(b)}
\end{overpic}
\caption{Normalized figure of merit (FOM) for the first and second harmonic  as a function of the duty cycle of the synchronous pump beam. The FOMs are shown for aligned and oriented states for (a) $\Omega_\textrm{dc}=0.05~\omega_\textrm{rf}$ and (b) $\Omega_\textrm{dc}=0.03~\omega_\textrm{rf}$. The parameters used for the calculation are the same as those in Fig.~\ref{fig:profiles}.}
\label{fig:duty}
\end{figure}

To optimize the vector-OPM response to external fields, one of the parameters to characterize is the amplitude of harmonics $h_{X,Y}^{(n)}$  with respect to the duty cycle of the pump beam. 
Fig.~\ref{fig:duty}~(a) shows the result for the normalized figure of merit (FOM)\footnote{The normalized FOM is defined as the relation between the absolute value of the mode and the linewidth normalized by the maximum of the set of duty cycle values.} between the mode amplitude $A$ and line-width,  $\Gamma$, of the harmonics as a function of the duty cycle for the three different states analyzed above.
Clearly, the three states show the same behavior of the FOM with respect the duty cycle, considering the rf amplitude as $\Omega_\mathrm{rf}=0.05~\omega_\textrm{rf}$.
One can notice that below $d=10\%$ duty cycle the FOM tends to zero.  At $d=10\%$, the three signals reach an optimum point and for higher duty cycles the two curves show that the FOM decrease.
It is worth noting that changing the duty cycle, without changing the pulse amplitude, reproduces the conditions the typical experimental conditions where the acousto-optical modulator (AOM) is used to generate synchronous pulses. In both cases, the changing duty cycle not only changes the duration of the pump interaction with the atomic spins, but also the effective power.
Hence, for $d=0\%$ there is no state being prepared resulting in zero Voigt rotation since the atoms are in a thermal state which has zero birefringence. In the intermediate region of the duty cycle, $0<d<10\%$, there is not enough time to prepare the optimal state for it to interact with the static fields and so the precession is small resulting in low birefringence. Around $d=10\%$ the spin dynamics is at its optimum where the interplay between the input state and the driven rf-field result in a strong precession around the static field. Lastly, when $d>10\%$ the pump dominates the interaction with the spin inhibiting the spin precession around the static field.

A different situation is observed with a lower rf amplitude $\Omega_\textrm{rf}=0.03~\omega_\textrm{rf}$. Fig.~\ref{fig:duty}~(b) shows the same behavior of the three states, but in this case the first harmonic $h_{X,Y}^{(1)}$ finds an optimum FOM at $d\sim 6\%$, different from the second harmonic with its maximum is still at $d=10\%$. For higher duty cycles the FOM decays faster compared to the case with $\Omega_\textrm{rf}=0.05~\omega_\textrm{rf}$. Notice that the first harmonic sensitivity to transverse fields implies that the state precesses around a tilted static field going off in the $x,y$ plane. Therefore a slight difference in the optimum duty cycle between $h^{(1)}$ and $h^{(2)}$ relies on the fact that a weaker rf driving field requires a shorter pump time for the precession to occur going off in the x-y plane. Otherwise the pump dominates, maintaining the spin precession close to the $x,y$ plane, which results in a high amplitude of $h^{(2)}$, but a reduction of the sensitivity for $h^{(1)}$.

\begin{figure}[b!]
\begin{overpic}[width=8.6cm]{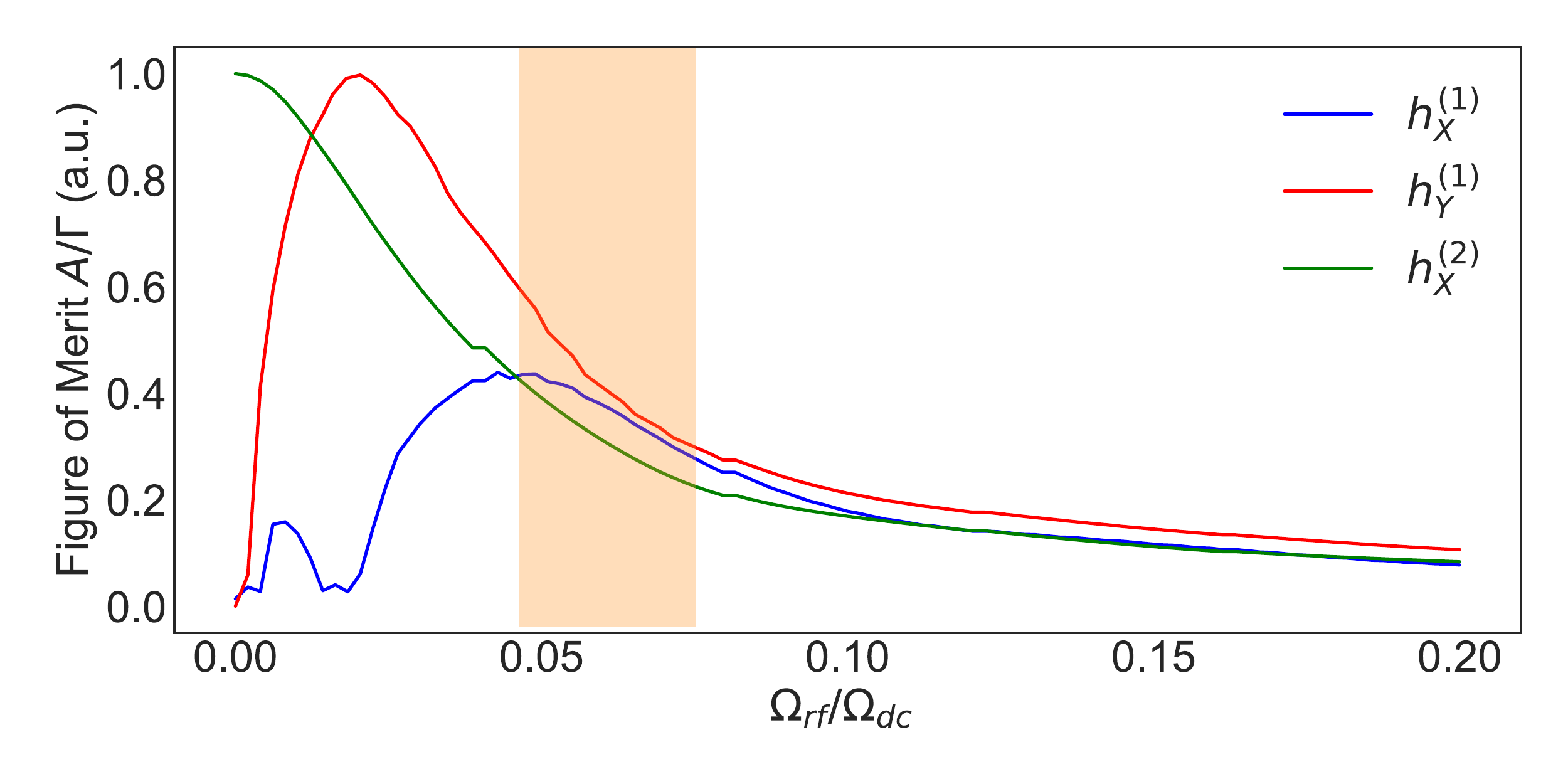}
\put (-5,50) {(a)}  
\end{overpic}
\begin{overpic}[width=0.5\textwidth]{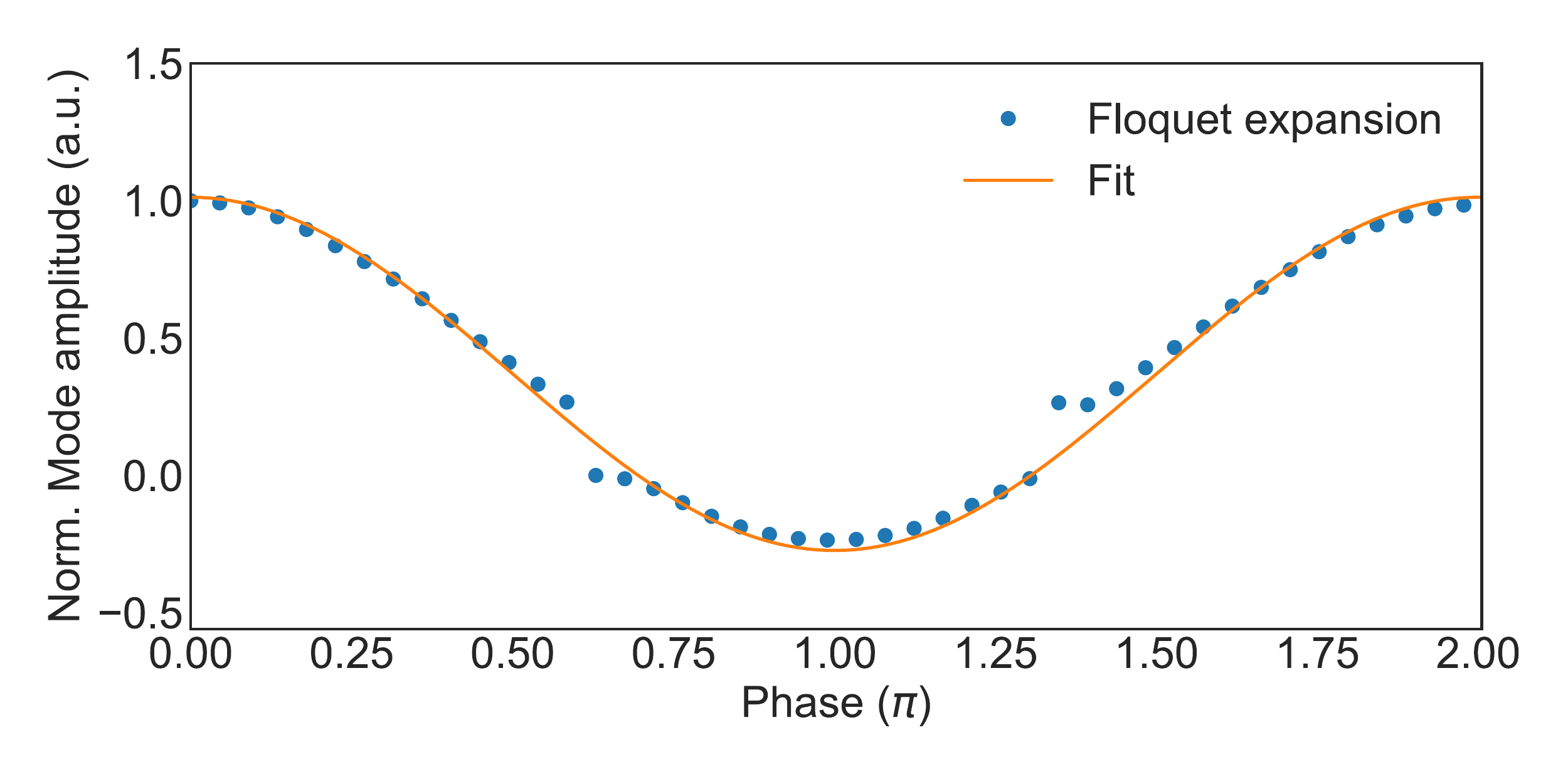}
\put (0,50) {(b)}
\end{overpic}
\caption{(a) FOM for $\omega$ and  $2\omega$ harmonics as a function of the radio-frequency dressing field amplitude $\Omega_{\textrm{rf}}$. The FOM for $h_X^{^{(1)}}$ and $h_Y^{^{(1)}}$ is normalized by the maximum of both signals, whereas the FOM for  $h_X^{^{(2)}}$ is normalized by its maximum on resonance. (b) Normalized mode amplitude of the second harmonic $h_X^{(2)}$ as a function of phase of the pump beam relative to the radio-frequency driving field. Here the pump is at 10\% duty cycle. The parameters used for the calculation are the same as those in Fig.~\ref{fig:profiles}.}
\label{fig:phase_and_amplitude}
\end{figure}

A second parameter that determines the optimisation of the vector-OPM, is the rf amplitude. Fig.~\ref{fig:phase_and_amplitude}~(a) shows the FOM as a function of the rf amplitude of the resonance profile of the first and second harmonic. This graph describes the behavior of the three states considered in the previous examples. A high FOM for the second harmonic $h_X^{(2)}$ is reached for very weak rf amplitudes, which corresponds to the narrowest magnetic resonance. By increasing the rf amplitude, the Voigt rotation signal at  $2\omega$ increases at the expense of broadening the magnetic resonance, which reduces the overall OPM sensitivity.
Meanwhile,  the maximum FOM for each quadrature $h_{X,Y}^{(1)}$ is reached for different values of rf amplitude. Whilst $h_{X}^{(1)}$ reaches its maximum at $\Omega_\mathrm{rf}\sim$ 0.025 $\Omega_\mathrm{dc}$, $h_{Y}^{(1)}$ reaches its maximum at $\Omega_\mathrm{rf}\sim$ 0.05 $\Omega_\mathrm{dc}$.
\begin{figure*}[t!]
\begin{overpic}[width=\textwidth]{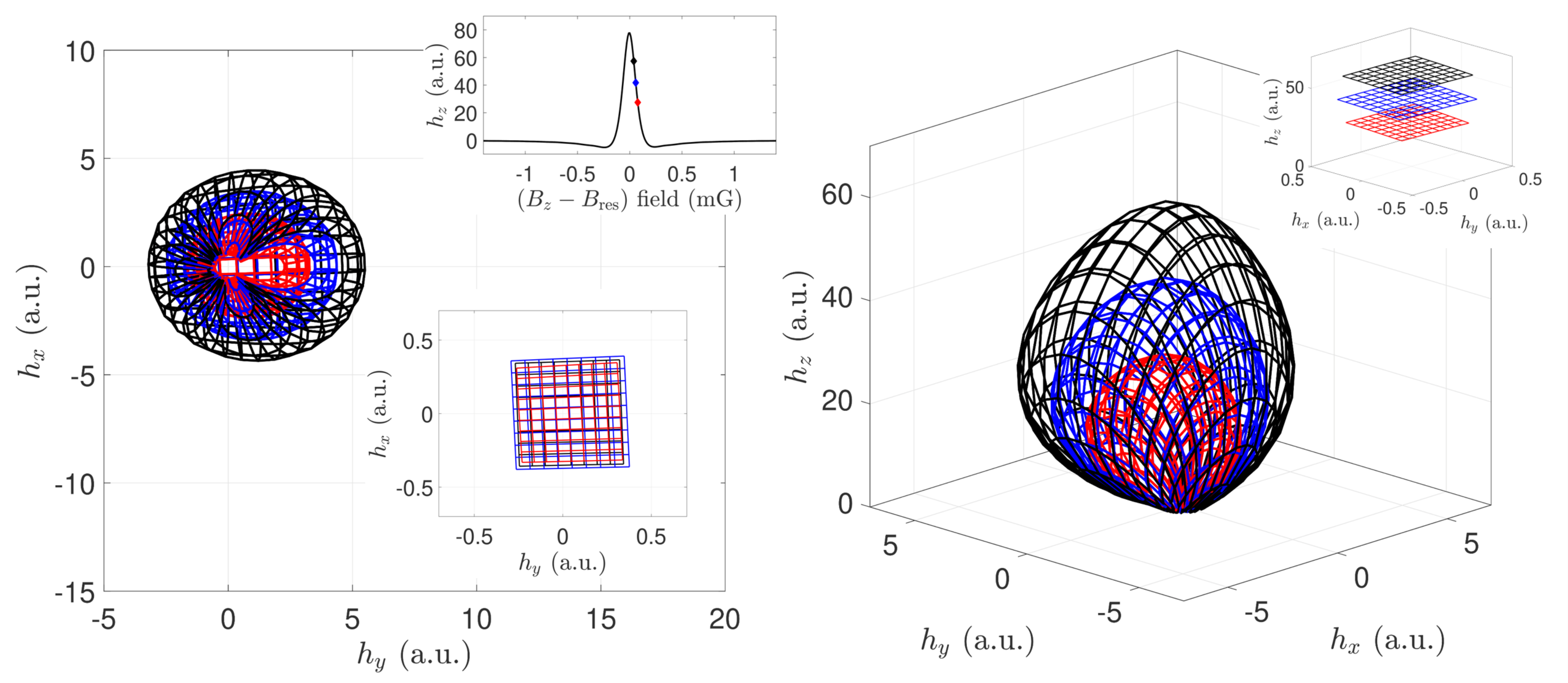}
\put (0,39) {(a)}  \put (53,39) {(b)}
\end{overpic}
\caption{Field mapping of  $h_x$, $h_y$ and $h_z$ with the external fields $B_x$, $B_y$ and $B_z$, considering the pump-probe sequence. Here, the atoms are dressed with a uniform 5~kHz rf field and pumped with a square intensity profile with a 10\% duty cycle, as in ref.~\cite{Tadas19}. 
The insets show the OPM response for small external fields, which behave linearly in the $h_x-h_y-h_z$ space. The parameters used for the calculation are the same as those in Fig.~\ref{fig:profiles}.  
}
\label{fig:3Dplot}
\end{figure*} 
For lower rf-amplitudes (below $\Omega_\textrm{rf}<0.045~\omega_\textrm{rf} $), we observe the highest relative difference between the quadratures, since the radio frequency field can not drive the tilted precession of the aligned state, whereas for higher amplitudes (above $\Omega_\textrm{rf}>0.075~\omega_\textrm{rf} $), the driving field broadens the resonance reducing the sensitivity~\footnote{
The FOM curve  is subjected to variances due to the algorithm that extract the maximum and minimum values to determine the FOM.}.
It can be observed that the optimization of the 3D operation is within the range near the maximum $\Omega_\mathrm{rf}\sim$ 0.05 $\Omega_\mathrm{dc}$ in which the three components are simultaneously sensitive to magnetic fields. Within this range   (light red square in Fig.~\ref{fig:phase_and_amplitude}~(a)) the three signals are non-zero and the relative FOM between the $h_{X,Y}^{(1)}$ do not exceed 80\%. This results in a comparable sensitivity between the external transverse and longitudinal fields.


In addition to the duty cycle and the rf amplitude optimization, Fig.~\ref{fig:duty}~(b) shows the normalised amplitude of the second harmonic $h_X^{(2)}$  a function of the pump phase with respect to the rf field. Notice the maximum amplitude of the second harmonic occurs in phase $(0,2\pi,\cdots)$, whereas a reduced and sign inverted amplitude is reached for anti-phase $(\pi,3\pi,\cdots)$. This curve describes the same behavior for the aligned and oriented states. The relative phase between the pump and the rf field strongly affects the state preparation process and the consequent OPM response to external fields.

\subsection{\label{subsec:detection}3D Vector mapping}

In this section we show how the theoretical model predicts the three-dimensional vector field mapping operation of the magnetometer. Following the same procedure as in ref.~\cite{Tadas19}, we can determine the three components of the field by setting the static field $B_z$ at $B^{3D}_z=B_\mathrm{res}+B_\mathrm{rf}/2$, which maximizes the mode amplitudes $h_{X,Y}^{(1)}$.
By linearly scanning the external transverse fields and demodulating the $h_{X,Y}^{(1)}$ and $h_{X}^{(2)}$ quadratures, we are able to map the magnetometer response.
In order to have a better picture of the spatial distribution of the external fields, we adopt a notation that relates the harmonics with the spatial directions $x,y$ and $z$ as 
\begin{align}
h_x=h_X^{(1)},~ h_y=h_Y^{(1)},~h_z=h_X^{(2)}.  
\end{align}
Figure~\ref{fig:3Dplot} shows the vector magnetometer operation visualized on a 3D plot for an stretched state $|2,\pm2\rangle_x$. Every oviform surface corresponds to the three different external longitudinal fields, $B^{\mathrm{ext}}_z$.
The Floquet expansion not only reproduces the oviform profile considering a large range of transverse fields as in ref.~\cite{Tadas19}, but also demonstrates the vector operation for small transverse fields with the three planes in the  $h_x-h_y-h_z$ space, which in the small field regime correspond to linear external field mapping 
\begin{align}
\lim_{B^\mathrm{ext}_x\ll B_\mathrm{dc}}h_x\propto B^\mathrm{ext}_x, \\ 
\lim_{B^\mathrm{ext}_y\ll B_\mathrm{dc}}h_y\propto B^\mathrm{ext}_y, \\ 
\lim_{B^\mathrm{ext}_z\ll B_\mathrm{dc}}h_z\propto B^\mathrm{ext}_z. 
\end{align}



One can observe in Fig.~\ref{fig:3Dplot}~(a) that with no transverse fields present, the spins are correctly mapped  at $h_x=h_y=0$. In this situation the aligned state precesses around the static field applied, perfectly aligned with the quantization axis of the of the probe beam, as discussed in sec.~\ref{sec:SecondMom_Aligned}. In the weak transverse field regime, the presence of orthogonal transverse fields $B_x$ and $B_y$  translates into the response of $h_x$ and $h_y$ quadratures as it is shown in insets of Figs.~\ref{fig:3Dplot}~(a) and (b).
At higher transverse fields, the resonant response is broader, inducing fast transitions among the Zeeman levels, leading to a thermal state for which the second harmonic is drastically reduced in which the three surfaces tend to $h_z=0$, e.g. $h_X^{(2)}\approx 0$ (see Fig.~\ref{fig:3Dplot}~(b)).

The same kind of vector-magnetometer response describes what is observed for the aligned state state $|F=2,m_F=0\rangle_x$ and the oriented state $|F=2,m_F=2\rangle_x$. The graphs are not shown independently to avoid repetition.

\section{\label{sec:Conclusions}Conclusions}

We have presented a theoretical model to describe the dynamics of a new kind of radio-frequency dressed three dimensional vector magnetometer based on the Voigt effect. As shown, our model describes the spin dynamics not only of the first moment, which is in agreement with the Bloch solution for Faraday based magnetometers, but also for the second moment.
We demonstrated that oriented and aligned states would present vector magnetometer response by dispersive Voigt rotation measurements, however, the aligned state is compatible with parallel geometry which is more suitable for miniature sensors.
In addition, we have shown that the time dependent dynamics involving synchronous pumping for state preparation can be solved employing a Floquet expansion, and the results are in agreement with the experimental observations in ref.~\cite{Tadas19}. The Floquet expansion is a powerful tool because it can solve different kinds of time dependent profiles making our approach general towards applications in understanding alternative approaches in OPM architecture. To determine the noise properties and sensitivity limits of the Voigt effect vector-OPM it would be necessary to solve the dynamics of the fourth moment of the spins. The model proposed in this work paves the way towards finding the solutions for higher moments. 
\\ \\

\section{Acknowledgements}
This work was funded by  Grant  No.  2018/03155-9  S\~ao  Paulo  Research  Foundation
(FAPESP) and by  Engineering and Physical Sciences
Research Council (EP/M013294/1). We thank Kasper Jensen for useful discussions and suggestions. 
\appendix

\section{Spin dynamics in the rotating frame \label{app:RotFrame}}
To show that eq.~(\ref{eq:FulldfdtLab2}) contains Bloch's solution for nuclear induction in the simplest case of constant input rate, we transform the dynamical equation into the rotating frame.

The time evolution of the spin operator is easily solved in a rotating frame, which oscillates at the same frequency as the rf field.
In the rotating frame, the atomic state is transformed as $\ket{\psi(t)'}=U(t)\ket{\psi(t)}$ whilst the spin operators follow $\hat{F}'_i(t)=U(t)^{-1}\hat{F}_i(t)U(t)$. Considering the unitary transformation $\hat{U}(t)=e^{i\omega t \hat{F}_z/\hbar}$,
the matrix representation of the rotating frame transformation around the $z$ axis is written as
 \begin{align}
\mathbf{\hat{F}}'(t)=\mathbf{R}(t)^{-1}\mathbf{\hat{F}}(t),\label{eq:dfdtRotframeApp}
\end{align}
where
\begin{align}
\mathbf{R}(t)&=\begin{bmatrix}
    \cos(\omega t) & \sin(\omega t) & 0 \\
    -\sin(\omega t) & \cos(\omega t) & 0 \\
    0 & 0 & 1
\end{bmatrix},\label{eq:RotMatrix}
\end{align}
and its inverse
\begin{align}
\mathbf{R}^{-1}(t)&=\begin{bmatrix}
    \cos(\omega t) & -\sin(\omega t) & 0 \\
    \sin(\omega t) & \cos(\omega t) & 0 \\
    0 & 0 & 1
\end{bmatrix}.
\end{align}
From eq.~(\ref{eq:dfdtRotframeApp}), we can obtain the dynamical equation in the rotating frame
\begin{align}
\frac{d\mathbf{\hat{F}}'(t)}{dt}=&\frac{d\mathbf{R}(t)^{-1}}{dt}\mathbf{R}(t) \mathbf{\hat{F}'}(t)+ \mathbf{R}(t)^{-1}\frac{d\mathbf{\hat{F}}(t)}{dt}. \label{eq:dftransformApp}
\end{align}
Considering the fact that the rotation matrix $ \mathbf{R}(t)$ can be expressed as 
 \begin{align}
\mathbf{R}(t)&=\mathbf{R}^{(0)} + \mathbf{R}^R \cos(\omega t) + \mathbf{R}^{I} \sin(\omega t),\\
\mathbf{R}^{-1}(t)&=\mathbf{R}^{(0)} + \mathbf{R}^R \cos(\omega t) - \mathbf{R}^{I} \sin(\omega t), \label{eq:RexpanssionApp}
\end{align}
where we have defined
\begin{align}
\mathbf{R}^R&=\begin{bmatrix}
    1 & 0 & 0 \\
    0 & 1 & 0 \\
    0 & 0 & 0
\end{bmatrix},\ 
\mathbf{R}^{I}=\begin{bmatrix}
    0 & -1 & 0 \\
    1 & 0 & 0 \\
    0 & 0 & 0
\end{bmatrix},\ 
\mathbf{R}^{(0)}&=\begin{bmatrix}
    0 & 0& 0 \\
   0 & 0 & 0 \\
    0 & 0 & 1
\end{bmatrix}.
\end{align}
which satisfy the following relations 
 \begin{align}
(\mathbf{R}^R)^2&=\mathbf{R}^R, (\mathbf{R}^{I})^2=-\mathbf{R}^R, \\
\mathbf{R}^R\mathbf{R}^{I} &=\mathbf{R}^{I}, \mathbf{R}^{I}\mathbf{R}^{R} =\mathbf{R}^{I},\\
\mathbf{R}^R \mathbf{R}^{(0)}&=0, \mathbf{R}^{I}\mathbf{R}^{(0)} =0.\label{eq:R1Rn1R0App}
\end{align}
From the relations above and the inverse matrix in  eq.~(\ref{eq:RexpanssionApp}), we obtain the following rotation rate
\begin{align}
\frac{d\mathbf{R}(t)^{-1}}{dt} = -\omega (\mathbf{R}^R \sin(\omega t) + \mathbf{R}^{I} \cos(\omega t)). \label{eq:dRdtApp}
\end{align}

Substituting eqs.~(\ref{eq:FulldfdtLabStoch}) and (\ref{eq:dRdtApp}) into eq.~(\ref{eq:dftransformApp}), we obtain the dynamics in the rotating frame
\begin{align}
\frac{d\mathbf{\hat{F}}'(t)}{dt}=& \frac{d\mathbf{R}(t)^{-1}}{dt}\mathbf{R}(t) \mathbf{\hat{F}}'(t)\nonumber\\
&+ \mathbf{R}(t)^{-1}[\tilde{\mathbf{B}}_0(t)+ \mathbf{B}^{\mathrm{ext}}]\mathbf{R}(t)\mathbf{\hat{F}}'(t)\nonumber\\
&- \Gamma_p(t)\mathbf{\hat{F}}'(t)+ \Gamma_p(t) \mathbf{R}(t)^{-1}\Braket{\mathbf{\hat{F}}^{\textrm{in}}}\nonumber\\
&+\mathbf{R}(t)^{-1}\bm{\mathcal{\hat{F}}}^{\textrm{in}}(t)+\mathbf{R}(t)^{-1}\bm{\mathcal{\hat{F}}}(t),\label{eq:dfdtrotApp}
\end{align}
where we have defined $\tilde{\mathbf{B}}_0(t)=\mathbf{B}_0(t)-\mathbf{\Gamma}_\mathrm{rel}$.
Expanding the first term on the right hand side and using the relations in eq.~(\ref{eq:R1Rn1R0App}) yields $\frac{d\mathbf{R}(t)^{-1}}{dt}\mathbf{R}(t) = -\omega \mathbf{R}^{I}$.
One can rewrite eq.~(\ref{eq:dfdtrotApp}) as
\begin{align}
\frac{d\mathbf{\hat{F}}'(t)}{dt}=
& (\mathbf{M}(t)+ \mathbf{M}^{\mathrm{ext}}(t))\mathbf{\hat{F}}'(t)- \Gamma_p(t)\mathbf{\hat{F}}'(t)\nonumber\\
& + \Gamma_p(t) \mathbf{R}(t)^{-1}\Braket{\mathbf{\hat{F}}^{\textrm{in}}}+{\bm{\mathcal{\hat{F}}}'}^{\textrm{in}}(t)+\bm{\mathcal{\hat{F}}}'(t),\label{eq:dfdtrot3}
\end{align}
where the stochastic operators transform as
${\bm{\mathcal{\hat{F}}}'}^{\textrm{in}}(t)=\mathbf{R}(t)^{-1}\bm{\mathcal{\hat{F}}}^{\textrm{in}}(t)$ and
$\bm{\mathcal{\hat{F}}}'(t)=\mathbf{R}(t)^{-1}\bm{\mathcal{\hat{F}}}(t)$, whereas the matrix $\mathbf{M}(t)$ is defined as 
\begin{widetext}
\begin{align}
\mathbf{M}(t)=& \mathbf{R}(t)^{-1}[\mathbf{B}_0(t)-\mathbf{\Gamma}_\mathrm{rel}]\mathbf{R}(t)-\omega  \mathbf{R}^{I} \\
=&\begin{bmatrix}
\frac{\Gamma_y \cos(2\omega t)}{2}-\frac{\Gamma_x \cos(2\omega t)}{2} -\left(\frac{\Gamma_x+\Gamma_y}{2}\right) & -\Delta +\frac{\Gamma_x \sin(2\omega t)}{2}-\frac{\Gamma_y \sin(2\omega t)}{2}& \Omega_{\mathrm{rf}}\sin(2\omega t)/2\\
    \Delta +\frac{\Gamma_x \sin(2\omega t)}{2}-\frac{\Gamma_y \sin(2\omega t)}{2} & -\frac{\Gamma_y \cos(2\omega t)}{2}+\frac{\Gamma_x \cos(2\omega t)}{2} -\left(\frac{\Gamma_x+\Gamma_y}{2}\right) & \frac{\Omega_{\mathrm{rf}}\cos(2\omega t)}{2}+ \frac{\Omega_{\mathrm{rf}}}{2} \\
    -\Omega_{\mathrm{rf}}\sin(2\omega t)/2 & -\frac{\Omega_{\mathrm{rf}}\cos(2\omega t)}{2}- \frac{\Omega_{\mathrm{rf}}}{2} & -\Gamma_z
\end{bmatrix},
\end{align}
\end{widetext}
where $\Delta=\Omega_{\mathrm{dc}}- \omega$ with $\Omega_{\mathrm{dc}}=\mu_B g'_F B_{\mathrm{dc}}$. The rotating frame transformation introduces terms oscillating at $2\omega$.
If we apply apply the rotating wave approximation (RWA) by neglecting those terms such that
\begin{align}
\mathbf{M}(t)=\mathbf{M}_{\mathrm{rot}} =&\begin{bmatrix}
 -\left(\frac{\Gamma_x+\Gamma_y}{2}\right) & -\Delta & 0\\
    \Delta &-\left(\frac{\Gamma_x+\Gamma_y}{2}\right) & \frac{\Omega_{\mathrm{rf}}}{2} \\
    0 & - \frac{\Omega_{\mathrm{rf}}}{2} & -\Gamma_z
\end{bmatrix}.\label{eq:M0App}
\end{align}
Under the RWA the dynamical matrix $\mathbf{M}(t)$ would be then time independent $\mathbf{M}(t)=\mathbf{M}_{\mathrm{rot}}$.

Regarding the contribution from external fields, the external matrix $\mathbf{B}^{\mathrm{ext}}(t)$  transforms into the matrix $\mathbf{M}^{\mathrm{ext}}(t)$  explicitly as 
\begin{align}
\mathbf{M}^{\mathrm{ext}}(t)=& \mathbf{R}(t)^{-1}\mathbf{B}^{\mathrm{ext}}\mathbf{R}(t)\\
=&\begin{bmatrix}
 0 & -\Omega_z^{\mathrm{ext}} & -\tilde{\Omega}_x(t)\\
  \Omega_z^{\mathrm{ext}}& 0 & \tilde{\Omega}_y(t) \\
    \tilde{\Omega}_x(t) & -\tilde{\Omega}_y(t) & 0
\end{bmatrix},\label{eq:MextApp}
\end{align}
where we have defined 
\begin{align}
\tilde{\Omega}_x(t) &=\Omega_x^{\mathrm{ext}}\sin(\omega t)-\Omega_y^{\mathrm{ext}}\cos(\omega t),\\
\tilde{\Omega}_y(t) &=\Omega_y^{\mathrm{ext}}\cos(\omega t)+\Omega_x^{\mathrm{ext}}\sin(\omega t).
\label{eq:eff_Omega_xy}
\end{align}

Hence, the generator of the dynamics can be defined as 
\begin{align}
\tilde{\mathbf{M}}(t)=\mathbf{M}_{\mathrm{rot}}+\mathbf{M}^{\mathrm{ext}}(t).    \label{eq:MtApp}
\end{align}

Therefore, the dynamics in the rotating frame is written as
\begin{align}
\frac{d\mathbf{\hat{F}}'(t)}{dt}=
& \tilde{\mathbf{M}}(t)\mathbf{\hat{F}}'(t)- \Gamma_p(t)\mathbf{\hat{F}}'(t)+ \Gamma_p(t) \mathbf{R}(t)^{-1}\Braket{\mathbf{\hat{F}}^{\textrm{in}}}\nonumber\\
& +{\bm{\mathcal{\hat{F}}}'}^{\textrm{in}}(t)+\bm{\mathcal{\hat{F}}}'(t),\label{eq:dfdtrot4}
\end{align}
where the stochastic operator transforms as $\bm{\mathcal{\hat{F}}}'(t)=\mathbf{R}\bm{\mathcal{\hat{F}}}(t)$ and the generator of the dynamics can be defined as 
$\tilde{\mathbf{M}}(t)=\mathbf{M}_{\mathrm{rot}}+\mathbf{M}^{\mathrm{ext}}(t)$
with 
\begin{align}
\mathbf{M}_{\mathrm{rot}} =&\begin{bmatrix}
 -\left(\frac{\Gamma_x+\Gamma_y}{2}\right) & -\Delta & 0\\
    \Delta &-\left(\frac{\Gamma_x+\Gamma_y}{2}\right) & \frac{\Omega_{\mathrm{rf}}}{2} \\
    0 & - \frac{\Omega_{\mathrm{rf}}}{2} & -\Gamma_z
\end{bmatrix},
\end{align}
according to eq.~(\ref{eq:M0App}) and  
\begin{align}
\mathbf{M}^{\mathrm{ext}}(t)=&\begin{bmatrix}
 0 & -\Omega_z^{\mathrm{ext}} & -\tilde{\Omega}_x(t)\\
  \Omega_z^{\mathrm{ext}}& 0 & \tilde{\Omega}_y(t) \\
    \tilde{\Omega}_x(t) & -\tilde{\Omega}_y(t) & -0
\end{bmatrix},
\end{align}
 according to eq.~(\ref{eq:MextApp}), where we have defined 
\begin{align}
\tilde{\Omega}_x(t) &=\Omega_x^{\mathrm{ext}}\sin(\omega t)-\Omega_y^{\mathrm{ext}}\cos(\omega t),\\
\tilde{\Omega}_y(t) &=\Omega_y^{\mathrm{ext}}\cos(\omega t)+\Omega_x^{\mathrm{ext}}\sin(\omega t).
\label{eq:eff_Omega_xy}
\end{align}

The generator $\tilde{\mathbf{M}}(t)$ is one of the most important matrices throughout the whole paper, since it contains all the interactions of the spin with the external magnetic fields and the relaxation terms.

At this point we can show that this result describes the magnetic resonance for polarized atomic spins. In this particular case, the relaxation matrix is $\Gamma_x=\Gamma_y=\Gamma_2$ and $\Gamma_z=\Gamma_1$
Additionally, we consider no external magnetic field i.e. $\mathbf{M}^{\mathrm{ext}}(t)=0$, an average pumping rate $\Gamma_p(t)=\Gamma_0$ and a polarized input state
$\langle\mathbf{\hat{F}}^{\textrm{in}}\rangle=(0,0,F_z^0)$, which is constant in the rotating frame $\mathbf{R}(t)\langle\mathbf{\hat{F}}^{\textrm{in}}\rangle=\langle\mathbf{\hat{F}}^{\textrm{in}}\rangle$. 
Hence, according to eq.~(\ref{eq:dfdtrot4}),
the steady state solution for the mean value is 
\begin{align}
& \Braket{\mathbf{\hat{F}}'(t)}=- (\mathbf{M}'_{\mathrm{rot}})^{-1}\Gamma_0 \Braket{\mathbf{\hat{F}}^{\textrm{in}}},
\end{align}
where the generator of the dynamics is
\begin{align}
\mathbf{M}'_{\mathrm{rot}} =&\begin{bmatrix}
  -\Gamma'_2 & -\Delta & 0\\
     \Delta &-\Gamma'_2 & \frac{\Omega_{\mathrm{rf}}}{2} \\
     0 & - \frac{\Omega_{\mathrm{rf}}}{2} & -\Gamma'_1
 \end{bmatrix},
 \end{align}
 defining $\Gamma'_i=\Gamma_i+\Gamma_0$ with $i=1,2$. This solution can be explicitly written in the Cartesian components as
\begin{align}
\Braket{\hat{F}'_x}&=\Braket{\hat{F}_z^{\textrm{in}}}\frac{\Gamma_0 \Delta \Omega'_{\mathrm{rf}}}{
(\Gamma'_2({\Omega_{\mathrm{rf}}'}^2+\Gamma'_1\Gamma'_2)+\Gamma'_1\Delta^2)},\\
\Braket{\hat{F}'_y}&=-\Braket{\hat{F}_z^{\textrm{in}}}\frac{\Gamma'_0 \Gamma'_2  \Omega'_{\mathrm{rf}}}{
(\Gamma'_2({\Omega_{\mathrm{rf}}'}^2+\Gamma'_1\Gamma'_2)+\Gamma'_1\Delta^2)},\\
\Braket{\hat{F}'_z}&=-\Braket{\hat{F}_z^{\textrm{in}}}\frac{( \Delta^2 +{\Gamma'_2}^2) \Gamma_0}{
(\Gamma'_2({\Omega_{\mathrm{rf}}'}^2+\Gamma'_1\Gamma'_2)+\Gamma'_1\Delta^2)},
\end{align}
where we have defined $\Omega'_{\mathrm{rf}}=\Omega_{\mathrm{rf}}/2$. This result
describes the magnetic resonance of a polarized atomic sample first obtained by Bloch \cite{Bloch46}.

\subsection{Floquet expansion for the first moment of the spin operator in the rotating frame \label{Sec:Floquete_1stFRot}}
According to eq.~(\ref{eq:dfdtrot4}) the dynamics for the mean value of of the spin operator are given by
\begin{align}
\frac{d\mathbf{P}'(t)}{dt}=
& (\mathbf{M}_{\mathrm{rot}}+ \mathbf{M}^{\mathrm{ext}}(t))\mathbf{P}'(t)- \Gamma_p(t)\mathbf{P}'(t)\nonumber\\
& + \Gamma_p(t) \mathbf{R}(t)^{-1}\Braket{\mathbf{P}^{\textrm{in}}},\label{eq:dSdtrot}
\end{align}
where the stochastic noise contribution is $\Braket{\bm{\mathcal{\hat{F}}}'(t)}=0$.
One of the interesting features of describing the dynamics in the rotating frame, is that unlike the laboratory frame where the external fields enter as a constant variable, in the rotating frame the transverse external fields naturally exhibit a contribution to the first harmonics.
In particular, the external field can be decomposed as 
 \begin{align}
\mathbf{M}^{\mathrm{ext}}(t)&=\mathbf{M}_{\mathrm{ext}}^{(0)}+ \mathbf{M}^{(1)} e^{i\omega t} + \mathbf{M}^{(-1)} e^{-i\omega t}, \label{eq:Mext}
\end{align}
where
\begin{align}
\mathbf{M}_{\mathrm{ext}}^{(0)}&=\begin{bmatrix}
    0 & -\Omega_z^{\mathrm{ext}} & 0 \\
    \Omega_z^{\mathrm{ext}} & 0 & 0 \\
   0& 0 & 0
\end{bmatrix},\\ 
\mathbf{M}^{(\pm 1)}&=\begin{bmatrix}
    0 & 0 & \pm i\Omega_- \\
    0 & 0 & \mp i\Omega_+ \\
   \mp i\Omega_-& \pm i\Omega_+ & 0
\end{bmatrix},\label{eq:M0pm1}
\end{align} 
with $\Omega_\pm=\Omega_x^{\mathrm{ext}}\pm i\Omega_y^{\mathrm{ext}}$. This result already shows that the transverse magnetic fields are mapped onto the quadrature of the spin evolution of the first harmonic $e^{i\omega t}$ since  $\Omega_x^{\mathrm{ext}}\pm i\Omega_y^{\mathrm{ext}}$. 

Similarly, the rotating frame matrix in eq.~(\ref{eq:RexpanssionApp})  can be decomposed as
 \begin{align}
\mathbf{R}^{-1}(t)&= \mathbf{R}^{(0)} + \mathbf{R}^{(1)} e^{i\omega t} + \mathbf{R}^{(-1)} e^{-i\omega t},\label{eq:RtN1}
\end{align}
where $\mathbf{R}^{(\pm 1)}=(\mathbf{R}^{R}\pm i \mathbf{R}^{I})/2$ with the definition of $\mathbf{R}^{R}$ and $\mathbf{R}^{I}$ given in the Appendix \ref{app:RotFrame}. This expansion implies that, according to the
pump rate decomposition in eq.~(\ref{eq:Ypdecomp}), the complete pumping term in eq.~(\ref{eq:dSdtrot}), can be expanded as
\begin{align}
\Gamma_p(t) \mathbf{R}(t)^{-1}=\sum \tilde{\mathbf{\Gamma}}_p^{(n)}e^{in\omega t},\label{eq:Ypdecomp2}
\end{align}
with 
\begin{align}
\tilde{\mathbf{\Gamma}}_p^{(n)}&=\mathbf{R}^{(0)} \Gamma_p^{(n)} +\mathbf{R}^{(1)} \Gamma_p^{(n+1)}+\mathbf{R}^{(-1)} \Gamma_p^{(n-1)}.
\end{align}

In order to solve this harmonic dynamical equation, we employ the Floquet expansion
for the spins in the rotating frame $\mathbf{P}'(t)=\sum_n\mathbf{P}'^{(n)}(t)e^{in\omega t}$.
Applying the same procedure in Sec.~\ref{sec:Floquet1rst_Lab},
we substitute the expansion $\mathbf{P}'(t)$ and $\mathbf{\Gamma}_p^{(n)}(t)$ into eq.~(\ref{eq:dSdtrot}) to obtain the the following dynamical equation for the vector of harmonics
\begin{align}
\frac{d\mathbb{P}'_F}{dt}=[\mathbb{M}-\mathbbm{\Gamma}]\ \mathbb{P}'_F +\tilde{\mathbbm{\Gamma}}_{\textrm{in}}\ \mathbb{S}_{\textrm{in}}, \label{eq:dSF_hiper_dtrot}
\end{align}
where
\begin{align}
\mathbb{M}_{nm}=\begin{cases}
      \mathbf{M}^{(0)}-in\omega \mathbf{I}_{3\times3}, & \text{for}\ n=m \\
      \mathbf{M}^{(\pm1)}, & \text{for}\ m=n\mp1 \\
      0, & \text{otherwise} \\
    \end{cases},
\end{align}
with $\mathbf{M}^{(0)}=\mathbf{M}_{\mathrm{rot}}+\mathbf{M}_{\mathrm{ext}}^{(0)}$ and the pump matrix term is 
$(\tilde{\mathbbm{\Gamma}}_{\textrm{in}})_{nm}= \tilde{\mathbf{\Gamma}}_p^{(n)}$ for $ n=m=0$, otherwise it is zero.

The steady state solution takes the same form as in the laboratory frame case, therefore,
\begin{align}
\mathbb{P}'_F=-[\mathbb{M}-\mathbbm{\Gamma}]^{-1}\tilde{\mathbbm{\Gamma}}_{\textrm{in}}\ \mathbb{P}_{\textrm{in}}. \label{eq:Sol_SF_hiper_rot}
\end{align}
However, the measurement of the spin dynamics is done in the laboratory frame. Therefore, applying the the inverse transformation $\mathbf{\hat{F}}(t)=\mathbf{R}(t)\mathbf{\hat{F}}'(t)$, the steady state solution can be expressed in the harmonic linear space as 
\begin{align}
\mathbb{P}_F=\mathbb{R}\ \mathbb{P}'_F=-\mathbb{R}\ [\mathbb{M}-\mathbbm{\Gamma}]^{-1}\tilde{\mathbbm{\Gamma}}_{\textrm{in}}\ \mathbb{S}_{\textrm{in}}, \label{eq:Sol_SF_hiper_rot_lab}
\end{align}
where
\begin{align}
\mathbb{R}_{nm}=\begin{cases}
      \mathbf{R}^{(0)}, & \text{for}\ n=m \\
      \mathbf{R}^{(\pm1)}, & \text{for}\ m=n\mp1 \\
      0, & \text{otherwise} \\
    \end{cases}.
\end{align}

\section{Diffusion matrix for spin operators \label{App:DiffusionM}}
The following description for Langevin dynamics is based in ref.~\cite{Julsgaard2003}. The relaxation dynamics of the spin operators under the effect of  stochastic operators can be written as 
\begin{align}
\frac{d\hat{F}_x(t)}{dt}&= - \Gamma_x\hat{F}_x(t)+\mathcal{\hat{F}}_x(t),\label{eq:dfxdt}\\
\frac{d\hat{F}_y(t)}{dt}&= - \Gamma_y\hat{F}_y(t)+\mathcal{\hat{F}}_y(t),\label{eq:dfydt}\\
\frac{d\hat{F}_z(t)}{dt}&= - \Gamma_z\hat{F}_z(t)+\mathcal{\hat{F}}_z(t).\label{eq:dfzdt}
\end{align}
We have defined in eq.~(\ref{eq:FFstochasticC}) that the stochastic operators $\mathcal{\hat{F}}_i(t)$ describe a white noise process, which satisfy the following correlation function 
\begin{align}
\Braket{\mathcal{\hat{F}}_i(t) \mathcal{\hat{F}}_j(t')}=\tilde{\Gamma}_{ij}\delta(t-t'),\label{eq:FFCorrFunctApp}
\end{align}
where $\Braket{\bm{\mathcal{\hat{F}}}(t)}=0$ and $\tilde{\Gamma}_{ij}=(\tilde{\Gamma})_{ji}$. To determine the elements of the diffusion matrix $\bm{\tilde{\Gamma}}$, we make use of the commutation relations of the spin operator $\hat{F}_i(t)$.

Let us start by solving the dynamical equation for the spin component $\mathcal{\hat{F}}_i(t)$ from eqs.~(\ref{eq:dfxdt}-\ref{eq:dfzdt}), such that 
\begin{align}
\hat{F}_i(t)&= e^{- \Gamma_i t}\hat{F}_i(0)+\int_0^t\mathcal{\hat{F}}_i(t)e^{\Gamma_i (t'-t)}dt'. \label{eq:fxt}
\end{align}

Now, this solution must satisfy the commutation relation $[\hat{F}_i(t),\hat{F}_j(t)]=i\epsilon_{ijk}\hat{F}_k(t)$ where $\epsilon_{ijk}$ is the Levi-Civita tensor,  for $t$ and $t+\Delta t$. Therefore, we can compute
\begin{widetext}
\begin{align}
\Braket{[\hat{F}_i(t+\Delta t),\hat{F}_j(t+\Delta t)]}=&e^{-(\Gamma_i+\Gamma_j)\Delta t}\Braket{[\hat{F}_i(t),\hat{F}_j(t)]}+e^{-(\Gamma_i+\Gamma_j)\Delta t}\int_0^{t+\Delta t}\int_0^{t+\Delta t}dt'dt''\nonumber\\
&\times e^{-\Gamma_i(t'-t)}e^{-\Gamma_j(t''-t)}\Braket{[\mathcal{\hat{F}}_i(t'),\mathcal{\hat{F}}_j(t'')]}.
\end{align}
\end{widetext}

From the correlation function of the stochastic operators in eq.~(\ref{eq:FFCorrFunctApp}) we obtain
\begin{align}
i\epsilon_{ijk} \Braket{\hat{F}_k(t+\Delta t)}&=i\epsilon_{ijk}e^{-(\Gamma_i+\Gamma_j)\Delta t}\Braket{\hat{F}_k(t)}\nonumber\\
&+\frac{(\tilde{\Gamma}_{ij}-\tilde{\Gamma}_{ji})}{\Gamma_i+\Gamma_j}(1-e^{-(\Gamma_i+\Gamma_j)\Delta t}).
\end{align}

In particular, is worth noting that for $\Delta t=0$ the spin operators have no change, and therefore satisfy the commutator relations.
Now, due to the decay process, for $\Delta t\ll1$  the spin operator is slightly attenuated and compensated by the diffusion matrix. To notice that, let us examine the influence of the diffusion matrix into the short time dynamics,  assuming $\Delta t\ll1$, such that at first order
\begin{align}
i\epsilon_{ijk} \Braket{\hat{F}_k(t+\Delta t)}&=i\epsilon_{ijk}[1-(\Gamma_i+\Gamma_j)\Delta t]\Braket{\hat{F}_k(t)}\nonumber\\
&+(\tilde{\Gamma}_{ij}-\tilde{\Gamma}_{ji})\Delta t.\label{eq:DifussionDyn}
\end{align}

If the diffusion matrix compensate the attenuation of the spin operator
we would obtain the following solution 
\begin{align}
\tilde{\Gamma}_{ij}-\tilde{\Gamma}_{ji}&=i(\Gamma_i+\Gamma_j)\epsilon_{ijk} \Braket{\hat{F}_k(t)}\\
&=(\Gamma_i+\Gamma_j)\Braket{[\hat{F}_i(t),\hat{F}_j(t)]},\label{eq:ComFiFj}
\end{align}
the spin operators would remain time independent, $\Braket{\hat{F}_k(t+\Delta t)}=\Braket{\hat{F}_k(t)}$.
The same procedure can be done for the anticommutator $\{\hat{F}_i(t),\hat{F}_j(t)\}$, such that 
\begin{align}
\tilde{\Gamma}_{ij}+\tilde{\Gamma}_{ji}&=(\Gamma_i+\Gamma_j)\Braket{\{\hat{F}_i(t),\hat{F}_j(t)\}}.\label{eq:AntiComFiFj}
\end{align}

In that particular case, from eqs.~(\ref{eq:ComFiFj}) and (\ref{eq:AntiComFiFj}), one can find that 
\begin{align}
\bm{\tilde{\Gamma}}=\mathbf{\Gamma}_\mathrm{rel}\ \bm{\sigma} +\bm{\sigma}\mathbf{\Gamma}_\mathrm{rel},
\end{align}
where we have defined the second moment matrix in the lab frame as $\bm{\sigma}=\Braket{\mathbf{\hat{F}}(t)\ \mathbf{\hat{F}}(t)^T}$. The explicit expression for the diffusion matrix is 
\begin{align}
\tilde{\mathbf{\Gamma}}
&=
\begin{bmatrix}
  2\Gamma_x\sigma_{xx} & (\Gamma_x+\Gamma_y)\sigma_{xy} & (\Gamma_x+\Gamma_z)\sigma_{xz} \\
  (\Gamma_x+\Gamma_y) \sigma_{yx} & 2\Gamma_y\sigma_{yy} & (\Gamma_y+\Gamma_z)\sigma_{yz} \\
    (\Gamma_x+\Gamma_z)\sigma_{zx} & (\Gamma_z+\Gamma_y)\sigma_{zy} & 2\Gamma_z\sigma_{zz}
\end{bmatrix}.
\end{align}

However, in general, the diffusion matrix does not necessarily compensate the attenuation of the spin operators, it can  reduce it though. Since the stochastic operators are modeling the flip in the atomic spins due to collision process, the second moment matrix can flip into the second moment of an unpolarized operator. Therefore, we can model the diffusion matrix as
\begin{align}
\tilde{\mathbf{\Gamma}}=\mathbf{\Gamma}_\mathrm{rel}\ \bm{\sigma}_0 +\bm{\sigma}_0\mathbf{\Gamma}_\mathrm{rel},\label{eq:Gammadifussion}
\end{align}
where $\bm{\sigma}_0$ is the second moment matrix of an unpolarized sample
that enters into the dynamics of the second moments of spin operators
at the rate of $\mathbf{\Gamma}_\mathrm{rel}$. Since $\bm{\sigma}_0$ satisfy the commutation relations, $\tilde{\mathbf{\Gamma}}$ guarantees that the spin operators satisfy the commutation relations as is given in eq.~(\ref{eq:DifussionDyn}).

The same procedure can be followed for the stochastic input operators $\bm{\mathcal{\hat{F}}}^{\textrm{in}}(t)$. Considering that the three directions are equally pumped, the diffusion rate takes a simpler form from eq.~(\ref{eq:Gammadifussion})
\begin{align}
\tilde{\mathbf{\Gamma}}=2\Gamma_p\ \bm{\sigma}_{\textrm{in}},\label{eq:GammaPdifussion}
\end{align}
where $\bm{\sigma}_{\textrm{in}}$ correspond to the second moment matrix for an arbitrary input state.

\subsection{Cross correlation functions \label{app:cross_correlations}}
Let us consider a stochastic operator $\bm{\mathcal{\hat{W}}}(t)$, where its mean value is
$\Braket{\bm{\mathcal{\hat{W}}}(t)}=0$ and the correlation function is 
\begin{align}
\Braket{\bm{\mathcal{\hat{W}}}(t) \bm{\mathcal{\hat{W}}}(t')^T}=\mathbf{\Gamma}_w(t)\ \delta(t-t'),\label{eq:CorrFunctLab}
\end{align}
The time evolution of a symmetric correlation is
\begin{align}
\bm{\mathcal{D}}_w=
& \int_{t_0}^t dt'\Braket{\bm{\mathcal{\hat{W}}}(t') \bm{\mathcal{\hat{W}}}(t)^T}+ \int_{t_0}^t dt'\Braket{\bm{\mathcal{\hat{W}}}(t) \bm{\mathcal{\hat{W}}}(t')^T}.\label{eq:WWstochT}
\end{align}
From the correlation function in eq.~(\ref{eq:CorrFunctLab})
we can write
\begin{align}
\bm{\mathcal{D}}_w=
& \int_{t_0}^t dt'\ \mathbf{\Gamma}_w(t)\delta(t-t')+ \int_{t_0}^t dt'\ \mathbf{\Gamma}_w(t')\delta(t'-t).\label{eq:D}
\end{align}
Using a change of variables, $t''=t-t'$ (and $t''=t'-t$ for the second integral) we obtain
\begin{align}
\bm{\mathcal{D}}_w=
& \int_{0}^{t-t_0} dt''\ \mathbf{\Gamma}_w(t)\delta(t'')+ \int_{-(t-t_0)}^0 dt''\ \mathbf{\Gamma}_w(t''+t)\delta(t'').\label{eq:D_2}
\end{align}
In the case of a time independent diffusion matrix $\mathbf{\Gamma}_w(t)=\mathbf{\Gamma}_{w_0}$
we find
\begin{align}
\bm{\mathcal{D}}_w=
& \mathbf{\Gamma}_{w_0},\label{eq:D_3}
\end{align}
which is in agreement with ref.~\cite{Cohen04}.
However for a time dependent diffusion matrix it is convenient
to apply an $\epsilon>0$ around zero, 
for a proper definition of the integral with a $\delta(t)$ function,
such that
\begin{align}
\bm{\mathcal{D}}_w=
& 2\mathbf{\Gamma}_w(t).\label{eq:D_4}
\end{align}

The cross correlation of the spin operator with stochastic operators in eq.~(\ref{eq:dsigma_dtLab2}) is given in a general form as 
\begin{align}
\bm{\mathcal{D}}=\Braket{\bm{\mathcal{W}}(t) \mathbf{\hat{F}}(t)^T}+
\Braket{\mathbf{\hat{F}}(t) \bm{\mathcal{W}}(t)^T},
\end{align}
which generates a drift in the second moment dynamics proportional to the diffusion matrix $\mathbf{\Gamma}_w(t)$. According to ref.~\cite{Cohen04}, this matrix is non-zero because $\mathbf{\hat{F}}(t)$ depends on the stochastic operators itself, therefore, the correlation $\Braket{\bm{\mathcal{W}}(t) \mathbf{\hat{F}}(t')^T}\neq0$ when $t'=t$, otherwise,
there is no correlation since $\bm{\mathcal{W}}(t)$ has a very short coherence time.

To determine the cross correlations $\bm{\mathcal{D}}$ with either stochastic operators, $\bm{\mathcal{W}}(t)\in \{\bm{\mathcal{\hat{F}}}^{\textrm{in}}(t),\bm{\mathcal{\hat{F}}}(t)\}$,
let us consider from eq.~(\ref{eq:FulldfdtLabStoch}), the time evolution of the spin operator as follows
\begin{align}
\mathbf{\hat{F}}(t)=&\mathbf{\hat{F}}(t_0)
+\int_{t_0}^tdt' [\mathbf{B}(t')+\mathbf{B}_{\mathrm{ext}}{(0)}-\Gamma_p(t')]\mathbf{\hat{F}}(t')\nonumber\\
& + \int_{t_0}^t dt'\left[\Gamma_p(t')\Braket{
\mathbf{\hat{F}}^{\textrm{in}}}+\bm{\mathcal{\hat{F}}}^{\textrm{in}}(t')\right]
+ \int_{t_0}^t dt'\bm{\mathcal{\hat{F}}}(t').\label{eq:IntF2}
\end{align}

In the case where $\bm{\mathcal{W}}(t)=\bm{\mathcal{\hat{F}}}^{\textrm{in}}(t)$, by multiplying from the right with $\bm{\mathcal{\hat{F}}}(t)^T$ and taking the average we have
\begin{align}
\Braket{\mathbf{\hat{F}}(t) \bm{\mathcal{\hat{F}}}^{\textrm{in}}(t)^T}=&\Braket{\mathbf{\hat{F}}(t_0) \bm{\mathcal{\hat{F}}}^{\textrm{in}}(t)^T}\nonumber\\
&+\int_{t_0}^tdt' \bar{\mathbf{B}}(t')\Braket{\mathbf{\hat{F}}(t') \bm{\mathcal{\hat{F}}}^{\textrm{in}}(t)^T}\nonumber\\
& + \int_{t_0}^t dt'\Gamma_p(t') \Braket{\mathbf{\hat{F}}^{\textrm{in}}}\Braket{ \bm{\mathcal{\hat{F}}}^{\textrm{in}}(t)^T}\nonumber\\
& + \int_{t_0}^t dt' \Braket{\bm{\mathcal{\hat{F}}}^{\textrm{in}}(t') \bm{\mathcal{\hat{F}}}^{\textrm{in}}(t)^T}\nonumber\\
&+ \int_{t_0}^t dt'\Braket{\bm{\mathcal{\hat{F}}}(t') \bm{\mathcal{\hat{F}}}^{\textrm{in}}(t)^T}.\label{eq:FFstochLab2}
\end{align}

It is worth pointing out that regardless the spin polarization $\Braket{\mathbf{\hat{F}}^{\textrm{in}}}$, the third term is zero, since
by definition $\Braket{ \bm{\mathcal{\hat{F}}}^{\textrm{in}}(t)^T}=0$. Moreover,
these kind of stochastic operators are only correlated with themselves, such that the correlation with any other operator $\mathbf{\hat{O}}(t)$ is $\Braket{\mathbf{\hat{O}}(t') \bm{\mathcal{\hat{F}}}^{\textrm{in}}(t)^T}=\Braket{\mathbf{\hat{O}}(t')}\Braket{ \bm{\mathcal{\hat{F}}}^{\textrm{in}}(t)^T}$, which by definition is zero.
Therefore, the only non-zero term is
\begin{align}
\Braket{\mathbf{\hat{F}}(t) \bm{\mathcal{\hat{F}}}^{\textrm{in}}(t)^T}\approx& \int_{t_0}^t dt' \Braket{\bm{\mathcal{\hat{F}}}^{\textrm{in}}(t') \bm{\mathcal{\hat{F}}}^{\textrm{in}}(t)^T}.\label{eq:FFstochLab3}
\end{align}
Following the same procedure for $\Braket{\bm{\mathcal{\hat{F}}}^{\textrm{in}}(t) \mathbf{\hat{F}}^(t)^T}$, and considering the case in eq.~(\ref{eq:D_4})
for the diffusion matrix in eq.~(\ref{eq:GammaPdifussion}), the cross correlation with the input operator is
\begin{align}
\Braket{\bm{\mathcal{\hat{F}}}^{\textrm{in}}(t)\ \mathbf{\hat{F}}(t)^T}+\Braket{\mathbf{\hat{F}}(t)\  \bm{\mathcal{\hat{F}}}^{\textrm{in}}(t)^T}=2\Gamma_p(t) \bm{\sigma}_{\textrm{in}}.\label{eq:FFstochLab4}
\end{align}

Similarly for the unpolarized stochastic operator $\bm{\mathcal{\hat{F}}}(t)$, the cross correlation is
\begin{align}
\Braket{\mathbf{\hat{F}}(t) \bm{\mathcal{\hat{F}}}(t)^T}=&\Braket{\mathbf{\hat{F}}(t_0) \bm{\mathcal{\hat{F}}}(t)^T}\nonumber\\
&+\int_{t_0}^tdt' \bar{\mathbf{B}}(t')\Braket{\mathbf{\hat{F}}(t') \bm{\mathcal{\hat{F}}}(t)^T}\nonumber\\
& + \int_{t_0}^t dt'\Braket{\mathbf{\hat{F}}^{\textrm{in}}(t') \bm{\mathcal{\hat{F}}}(t)^T}\nonumber\\
&+ \int_{t_0}^t dt'\Braket{\bm{\mathcal{\hat{F}}}(t') \bm{\mathcal{\hat{F}}}(t)^T},\label{eq:FFstochLab5}
\end{align}
where the only non-zero term is
\begin{align}
\Braket{\mathbf{\hat{F}}(t) \bm{\mathcal{\hat{F}}}(t)^T}=& \int_{t_0}^t dt'\Braket{\bm{\mathcal{\hat{F}}}(t') \bm{\mathcal{\hat{F}}}(t)^T}.\label{eq:FFstochLab6}
\end{align}

Since the drift term $\tilde{\mathbf{\Gamma}}$ in eq.~(\ref{eq:FFstochasticC}) is constant, according to eq.~(\ref{eq:D_3}), the complete cross correlation is
\begin{align}
\Braket{\mathbf{\hat{F}}(t) \bm{\mathcal{\hat{F}}}(t)^T}+\Braket{\bm{\mathcal{\hat{F}}}(t) \mathbf{\hat{F}}(t)^T}=&\  \tilde{\mathbf{\Gamma}}.\label{eq:FFstochLab6}
\end{align}
From eq.~(\ref{eq:Gammadifussion}), we find that the drift matrix of the stochastic operator can take the form
\begin{align}
\tilde{\mathbf{\Gamma}}=\mathbf{\Gamma}_\mathrm{rel}\ \bm{\sigma}_{0}+\bm{\sigma}_{0}\ \mathbf{\Gamma}_\mathrm{rel},\label{eq:FFstochLab7}
\end{align}
where $\bm{\sigma}_0$ represents the second moment matrix
of an unpolarized or thermal spin state.

\section{Rotating Frame transformation of the second moment matrix dynamics
\label{app:sigma_rot_transform}}

Taking the time derivative of eq.~(\ref{eq:sigmarot}) we have
\begin{align}
\frac{d\bm{\sigma}'(t)}{dt}=& \frac{d\mathbf{R}^{-1}(t)}{dt}\mathbf{R}(t) \bm{\sigma}'(t)+ \mathbf{R}^{-1}(t)\frac{d\bm{\sigma}(t)}{dt}\mathbf{R}(t)\nonumber\\
& \bm{\sigma}'(t)\mathbf{R}^{-1}(t)\frac{d\mathbf{R}(t)}{dt}, \label{eq:dsigmadtrotApp}
\end{align}
and from eq.~(\ref{eq:dRdtApp}) we obtain
\begin{align}
\frac{d\bm{\sigma}'(t)}{dt}=& \omega (\mathbf{R}^{I} \bm{\sigma}'(t)-\bm{\sigma}'(t)\mathbf{R}^{I} ) + \mathbf{R}^{-1}(t)\frac{d\bm{\sigma}(t)}{dt}\mathbf{R}(t).
\label{eq:dsigmadtrotApp2}
\end{align}

Substituting eq.~(\ref{eq:dsigma_dtLabPol}) into the rotating frame dynamics, considering the rotating wave approximation, leads to 
\begin{align}
\frac{d\bm{\sigma}(t)}{dt}=&\tilde{\mathbf{M}}(t) \bm{\sigma}'(t)+\bm{\sigma}'(t)\tilde{\mathbf{M}}(t)^T
-2\Gamma_p(t)\bm{\sigma}'(t)\nonumber\\
&+\mathbf{\Gamma}'_\mathrm{rel}\ \bm{\sigma}'_{0}+\bm{\sigma}'_{0}\ \mathbf{\Gamma}'_\mathrm{rel}+2\Gamma_p(t)\ \bm{\sigma}'_{\textrm{in}}(t)\nonumber\\
&+\Gamma_p(t)\left[\mathbf{R}(t)^{-1}\Braket{\mathbf{\hat{F}}^{\textrm{in}}} \Braket{\mathbf{\hat{F}}'(t)^T}\right.\nonumber\\
&\hspace{2cm}\left.+\Braket{\mathbf{\hat{F}}'(t)} \Braket{\mathbf{\hat{F}}^{\textrm{in}~T}}\mathbf{R}(t)\right],\label{eq:dsigmadtrotApp3}
\end{align}
in which $\tilde{\mathbf{M}}(t)$ is given in eq.~(\ref{eq:dfdtrot4}),
the unpolarized second moment matrix remains diagonal
\begin{align}
\bm{\sigma}'_{0}&=\mathbf{R}(t)^{-1}\bm{\sigma}_{0}\mathbf{R}(t)=\bm{\sigma}_{0},
\end{align}
and the relaxation matrix is such that
\begin{align}
\mathbf{\Gamma}'_\mathrm{rel}&=\mathbf{R}(t)^{-1}\mathbf{\Gamma}_\mathrm{rel}\mathbf{R}(t)\nonumber\\
&=\begin{bmatrix}
    \frac{\Gamma_x+\Gamma_y}{2} & 0 &0  \\
    0 & \frac{\Gamma_x+\Gamma_y}{2} & 0 \\
    0 & 0 & \Gamma_z
\end{bmatrix}.\label{eq:Yrelapp}
\end{align}

In addition, the input second moment matrix is now time dependent $\bm{\sigma}'_{\textrm{in}}(t)=\mathbf{R}(t)^{-1}\bm{\sigma}_{\textrm{in}}\mathbf{R}(t)$ 
 \begin{align}
 \bm{\sigma}'_{\textrm{in}}(t)&={\bm{\sigma}'_{\textrm{in}}}^{(0)} + {\bm{\sigma}'_{\textrm{in}}}^{(1)}e^{i\omega t} +{\bm{\sigma}'_{\textrm{in}}}^{(-1)}e^{-i\omega t},\label{eq:sigma_in_rot}
\end{align}
where
 \begin{align}
 {\bm{\sigma}'_{\textrm{in}}}^{(0)}=&\mathbf{R}^{(0)}\ \bm{\sigma}_{\textrm{in}}\ \mathbf{R}^{(0)}\nonumber\\
&+\mathbf{R}^{(1)}\ \bm{\sigma}_{\textrm{in}}\ \mathbf{R}^{(1)}
+\mathbf{R}^{(-1)}\ \bm{\sigma}_{\textrm{in}}\ \mathbf{R}^{(-1)}\\
{\bm{\sigma}'_{\textrm{in}}}^{(\pm1)}=&[\mathbf{R}^{(0)}\ \bm{\sigma}_{\textrm{in}}\ \mathbf{R}^{(\mp1)} +\mathbf{R}^{(\pm 1)}\ \bm{\sigma}_{\textrm{in}} \ \mathbf{R}^{(0)}].
\end{align}

Considering a non-polarized input state $\Braket{\mathbf{\hat{F}}^{\textrm{in}}}=0$, we can rewrite eq.~(\ref{eq:dsigmadtrotApp3}) as 
\begin{align}
\frac{d\bm{\sigma}(t)}{dt}=&\tilde{\mathbf{M}}(t) \bm{\sigma}'(t)+\bm{\sigma}'(t)\tilde{\mathbf{M}}(t)^T
-2\Gamma_p(t)\bm{\sigma}'(t)\nonumber\\
&+\mathbf{\Gamma}'_\mathrm{rel}\ \bm{\sigma}_{0}+\bm{\sigma}_{0}\ \mathbf{\Gamma}'_{rel}+2\Gamma_p(t)\ \bm{\sigma}'_{\textrm{in}}(t).\label{eq:dsigmadtrotApp4}
\end{align}

In the Liouville space, the dynamics equations for the second moment is given by eq.~(\ref{eq:dsigma_dt_LioRot}). In that case, the transformation of the input second moment matrix can be written as
$\mathbf{X}'_{\textrm{in}}(t)=\mathsf{R}_{\textrm{in}}(t)\mathbf{X}_{\textrm{in}}$ with
\begin{align}
\mathsf{R}_{\textrm{in}}(t)=&\mathsf{R}_{\textrm{in}}^{(0)}+\mathsf{R}_{\textrm{in}}^{(1)}e^{i\omega t}+\mathsf{R}_{\textrm{in}}^{(-1)}e^{-i\omega t},\label{eq:Rin}
\end{align}
where $\mathsf{R}_{\textrm{in}}^{(0)}=\mathsf{R}^{(0)}$, $\mathsf{R}_{\textrm{in}}^{(\pm1)}=\mathsf{R}^{(\mp1)}$, which are defined in eqs.~(\ref{eq:RLR0}) and (\ref{eq:RLR1}).
In the case where this transformation matrix is modulate by the pump rate, $\Gamma_p(t)\mathsf{R}_{in}(t)$, the effective time dependence can be harmonically expanded as $\Gamma_{p}'(t)=\Gamma_p(t)\mathsf{R}_{in}(t)=\sum_n {\mathbf{\Gamma}'_{p}}^{(n)}e^{in\omega t}$ where
\begin{align}
{\mathbf{\Gamma}'_{p}}^{(n)}=&
\mathsf{R}^{(0)}\Gamma_p^{(n)}+\mathsf{R}^{(1)}\Gamma_p^{(n-1)}+\mathsf{R}^{(-1)}\Gamma_p^{(n+1)}.\label{eq:gammaRot_n}
\end{align}

Therefore, the dynamics in the Liouville space  is given by 
\begin{align}
\frac{d\mathbf{X}'(t)}{dt}=&\mathbf{G}(t)\mathbf{X}'(t)-2\Gamma_p(t)\ [ \mathbf{X}'(t)-\mathsf{R}_{\textrm{in}}(t)\mathbf{X}_{\textrm{in}}]
+\Lambda'_\mathrm{rel}\ \mathbf{X}_{0},\label{app:dsigma_dt_LioRot}
\end{align}

\subsection{Transformation of the second moment matrix to the laboratory frame \label{app:RotTrans_sigma}}
The rotating frame transformation for $\bm{\sigma}$ is 

 \begin{align}
\bm{\sigma}(t)&=\mathbf{R}(t)\ \bm{\sigma}'(t)\  \mathbf{R}^{-1}(t).\label{eq:LabTransf}
\end{align}

Therefore, in the Liouville space, we can rewrite the transformation as $\mathbf{X}(t) =\mathsf{R}(t)\mathbf{X}'(t)$ where $\mathsf{R}=\mathcal{L}(\mathbf{R}(t))\mathcal{R}(\mathbf{R}^{-1}(t))$.
Hence in terms of harmonics we have
%
 \begin{align}
\mathsf{R}(t)=&[\mathsf{R}^{(0)}+\mathsf{R}^{(1)}e^{i\omega t}+\mathsf{R}^{(-1)}e^{-i\omega t}\nonumber\\
&+\mathsf{R}^{(2)}e^{2i\omega t} +\mathsf{R}^{(-2)}e^{-2i\omega t}],\label{eq:RLab2}
\end{align}
with
 \begin{align}
\mathsf{R}^{(0)}&=
\left[\mathsf{R}_L^{(0)}\mathsf{R}_R^{(0)}+\mathsf{R}_L^{(1)}\mathsf{R}_R^{(1)} +\mathsf{R}_L^{(-1)}\mathsf{R}_R^{(-1)}\right],\label{eq:RLR0}\\
\mathsf{R}^{(\pm 1)}&=\left[\mathsf{R}_L^{(0)}\mathsf{R}_R^{(\pm 1)}+\mathsf{R}_L^{(\mp1)}\mathsf{R}_R^{(0)}\right],\label{eq:RLR1}\\
\mathsf{R}^{(\pm 2)}&=\mathsf{R}_L^{(\mp1)}\mathsf{R}_R^{(\pm1)},\label{eq:RLR2}
\end{align}
where we have defined $\mathsf{R}_L^{(n)}=\mathcal{L}(\mathbf{R}^{(n)})$ and $\mathsf{R}_R^{(n)}=\mathcal{R}(\mathbf{R}^{(n)})$.

%
\end{document}